\documentclass[submitting]{nst}
\usepackage[T2A,T1]{fontenc}
\usepackage[russian,english]{babel}
\usepackage{indentfirst}
\usepackage{subcaption}
\usepackage{graphicx}
\usepackage{multirow}
\usepackage{placeins}
\usepackage{makecell}
\usepackage{titlesec}  
\usepackage{float}
\usepackage{booktabs}
\usepackage{amsmath}
\usepackage{pifont}
\usepackage{mathtools}  
\usepackage{dcolumn}
\usepackage{caption}
\usepackage{textcomp}
\usepackage{listings}
\usepackage{siunitx}
\DeclareSIUnit{\standardliter}{SL}
\DeclareSIUnit{\ppb}{ppb}
\usepackage{microtype} 
\usepackage[none]{hyphenat}
\nolinenumbers
\renewcommand{\linenumbers}{}

\makeatletter
\providecommand{\inst}[1]{\@textsuperscript{\normalfont#1}}

\makeatother

\usepackage{caption}
\usepackage{ragged2e} 
\DeclareCaptionJustification{myjustify}{\justifying}
\newcommand{\tsinghuaphy}{\affiliation{Department of Physics \& Center for High Energy Physics, Tsinghua University, Beijing 100084, China}}
\newcommand{\tsinghuaephy}{\affiliation{Key Laboratory of Particle and Radiation Imaging (Ministry of Education) \& Department of Engineering Physics, Tsinghua University, Beijing 100084, China}}
\newcommand{\ustc}{\affiliation{State Key Laboratory of Particle Detection and Electronics, University of Science and Technology of China, Hefei 230026, China}}
\newcommand{\ustcimp}{\affiliation{
		Department of Modern Physics, University of Science and Technology of China, Hefei 230026, China}}
\newcommand{\westlake}{\affiliation{School of Science, Westlake University, Hangzhou 310030, China}}
\newcommand{\sysuphy}{\affiliation{School of Physics, Sun Yat-sen University, Guangzhou 510275, China}}
\newcommand{\sysuifcen}{\affiliation{Sino-French Institute of Nuclear Engineering and Technology, Sun Yat-sen University, Zhuhai 519082, China}}
\newcommand{\cuhk}{\affiliation{School of Science and Engineering, The Chinese University of Hong Kong (Shenzhen), Shenzhen, Guangdong, 518172, China}}
\newcommand{\buaasp}{\affiliation{School of Physics, Beihang University, Beijing 100083, China}}
\newcommand{\buaalab}{\affiliation{Beijing Key Laboratory of Advanced Nuclear Materials and Physics, Beihang University, Beijing 100191, China}}
\newcommand{\sanmen}{\affiliation{CNNC Sanmen Nuclear Power Company, Zhejiang 317112, China}}

\begin{document}

\author{Jia-Chen Yu}\ustc\ustcimp

\author{Kai-Hang Li}\tsinghuaphy

\author{Jing-Fan Gu}\tsinghuaphy
\author{Chang Cai}\tsinghuaphy
\author{Guo-Cai Chen}\sanmen
\author{Jiang-Yu Chen}\sysuifcen
\author{Hua-Yu Dai}\cuhk
\author{Run-Dong Fang}\buaasp
\author{Hong-Rui Gao}\tsinghuaphy
\author{Fei Gao}\tsinghuaphy
\author{Xiao-Ran Guo}\ustc\ustcimp
\author{Ji-Heng Guo}\buaasp
\author{Cheng-Jie Jia}\tsinghuaphy\thanks{Now at: Department of Physics, Stanford University, Stanford, CA 94305, USA}
\author{Gao-Jun Jin}\sanmen
\author{Fa-Li Ju}\sanmen
\author{Yan-Zhou Hao}\tsinghuaphy
\author{Xu Han}\tsinghuaphy
\author{Yang Lei}\tsinghuaphy
\author{Meng Li}\sanmen
\author{Min-Hua Li}\sanmen
\author{Sheng-Chao Li}\westlake
\author{Si-Yin Li}\westlake
\author{Tao Li}\sanmen
\author{Qing Lin}\thanks{\emph{Corresponding author: } qinglin@ustc.edu.cn} \ustc\ustcimp
\author{Jia-Jun Liu}\sysuphy
\author{Sheng Lv}\sanmen
\author{Yuan-Yuan Ren}\cuhk
\author{Chuan-Ping Shen}\sanmen
\author{Li-Jun Tong}\ustc\ustcimp
\author{Yu-Huang Wan}\tsinghuaphy
\author{Jun Wang}\westlake
\author{Xiao-Yu Wang}\westlake
\author{Wei Wang}\sysuphy\sysuifcen
\author{Xiao-Ping Wang}\buaasp\buaalab
\author{Zi-Hu Wang}\sanmen
\author{Yue-Huan Wei}\sysuifcen
\author{Li-Ming Weng}\sanmen
\author{Xiang Xiao}\sysuphy
\author{Ling-Feng Xie}\tsinghuaphy
\author{Ji-Jun Yang}\westlake
\author{Li-Tao Yang}\tsinghuaephy
\author{Long Yang}\sanmen
\author{Jing-Qiang Ye}\cuhk
\author{Qian Yue}\tsinghuaephy
\author{Yu-Yong Yue}\sysuifcen
\author{Tian-Yuan Zha}\cuhk
\author{Bing-Wei Zhang}\sanmen
\author{Hong-Hui Zhang}\cuhk
\author{Zhi-Cai Zhang}\tsinghuaephy
\author{Yi-Fei Zhao}\tsinghuaphy

\collaboration{RELICS Collaboration}
\thanks{\emph{Collaboration email: } relics@tsinghua.edu.cn}
\noaffiliation

\title{Projection of purification performance for the RELICS experiment}

\begin{abstract}
The RELICS (REactor neutrino LIquid xenon Coherent elastic Scattering) experiment employs a dual-phase liquid xenon time projection chamber to search for Coherent Elastic Neutrino-Nucleus Scattering (CE$\nu$NS) induced by reactor neutrinos. To detect these sub-keV nuclear recoils and minimize signal attenuation, it is critical to maintain a sufficiently low impurity concentration in the detector. This work presents a comprehensive purity evolution model developed to describe impurity migration inside the detector. Utilizing measured material outgassing rates as input parameters, the model incorporates non-uniform transport mechanisms of the impurities, including circulation, vaporization, and condensation. The model is validated using data from a dedicated prototype detector. Based on this validated model, projections for the purification performance of the upcoming RELICS-10 and RELICS-50 detectors are provided.
\end{abstract}


\maketitle

\section{Introduction}\label{sec:sec}

Coherent elastic neutrino-nucleus scattering (CE$\nu$NS) was first predicted in 1974 within the Standard Model framework, where a neutrino scatters off an entire nucleus with a coherent cross-section scaling with the square of the neutron number ($N^2$) at low momentum transfer~\cite{cevns_theory,kopeliovich1974isotopic}.
This process serves as a precision probe for the electroweak sector~\cite{canas2018future}, astrophysical dynamics~\cite{beacom2010diffuse}, and Beyond Standard Model physics such as non-standard interactions and sterile neutrinos~\cite{Coloma:2017egw,PhysRevC.95.025801}. Following the breakthrough observation of CE$\nu$NS by the COHERENT collaboration using spallation neutrons, which has been subsequently confirmed in CsI(Na), liquid argon and germanium detectors~\cite{COHERENT:2017ipa,akimov2021first,adamski2025evidence}, signals have now been detected across a broad energy range. This includes the detection of solar neutrinos in multi-ton liquid xenon dark matter experiments, as recently reported by the XENONnT and PandaX-4T collaborations~\cite{PhysRevLett.133.191002,PhysRevLett.133.191001}. Additionally, the recent observation of reactor antineutrino CE$\nu$NS by CONUS+ on a germanium target represents the first \SI{3}{\sigma}-level signal from a reactor source~\cite{Ackermann:2025}. Despite these remarkable achievements, the detection of CE$\nu$NS signals at reactor sites remains a significant frontier. Achieving high-precision measurements requires further signal validation with different target materials and independent detection techniques to provide independent verification of the observation and systematically control uncertainties.

\section{The RELICS Experimental Program}\label{sec:design}

The RELICS (REactor neutrino LIquid xenon Coherent elastic Scattering) program follows a staged implementation plan to conduct tests at the Sanmen nuclear reactor site using dual-phase liquid xenon time projection chamber (LXeTPC) technology~\cite{tpc}, progressively validating detector technologies and achieving the sensitivity required for CE$\nu$NS observation~\cite{cai2024reactor,qian2019physics}. This roadmap comprises three phases. The initial phase utilizes a prototype detector to verify the performance of key techniques. Specifically, dedicated campaigns were conducted to investigate purification schemes and validate the purity evolution model. In the following stage, a detector with an active mass of \SI{10}{\kilo\gram}, designated as RELICS-10, will be operated as the second phase of the RELICS experiment aimed at validating long-term system stability at a reactor site. Its active volume is enclosed by a field cage lined with highly reflective polytetrafluoroethylene (PTFE) panels to optimize light collection, and a bottom overflow chamber is employed to ensure passive and stable liquid level control. Ultimately, the full-scale RELICS-50 detector, scaling the active xenon mass to approximately \SI{50}{\kilo\gram}, will be constructed and operated for high-precision physics searches. They will share a standardized framework for their cryogenic and purification systems.

This section is divided into three parts. The first part introduces the overall design of the circulation and purification systems for the RELICS experiment, including the TPC designs, as well as the various subsystems and circulation modes required for operation. Building upon these hardware descriptions, the second part details the specific detector designs employed in the Run~7 and Run~9 prototype campaigns (the seventh and ninth physics runs, respectively), along with the key experimental operations. These operations influence the purity evolution, thereby enabling the characterization of the purity evolution and providing a robust validation of the purity model. The third part describes how the data obtained from the detectors are used to evaluate the purity performance of the detector, as well as the relationship between purity and signal sensitivity.

\subsection{The RELICS Circulation and Purification System Design}

\label{sec:system design}
\begin{figure}[htbp]
    \centering
    \includegraphics[width=0.99\linewidth]{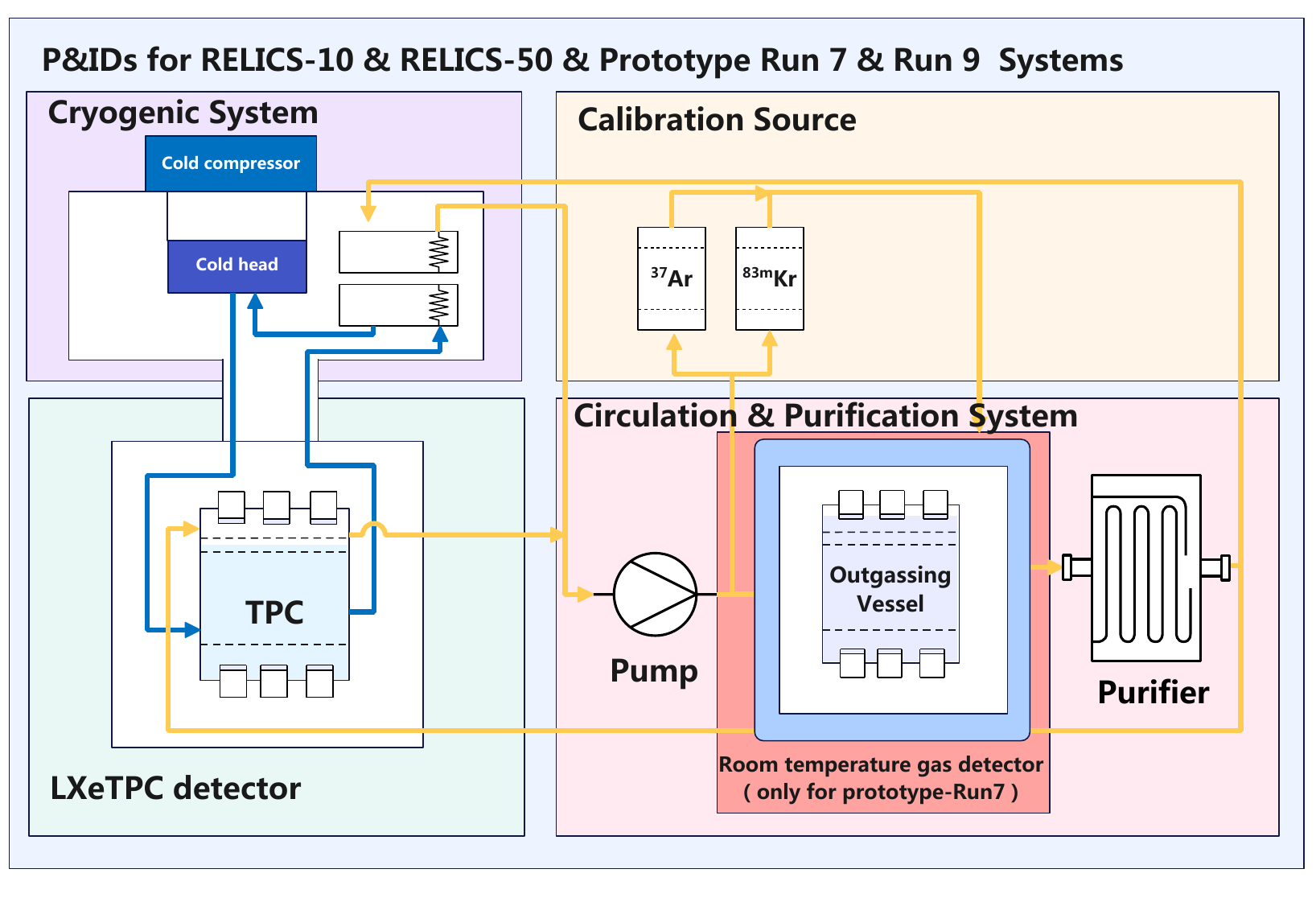}
    \caption{Piping and Instrumentation Diagram (P\&ID) of the RELICS system. The diagram shows the interconnection between the TPC, the cryogenic system, the calibration sources, and the circulation \& purification system. The blue and orange lines represent LXe and GXe flows, respectively, with arrows indicating the direction of circulation. Components such as the outgassing vessel and room-temperature gas detector were utilized specifically for the Run~7 prototype campaign.}
    \label{fig:relics_prototypes_PID}
\end{figure}

The experimental setup comprises several subsystems: the LXeTPC detector, the cryogenic system, the circulation and purification systems, and the calibration sources. Among these, the circulation and purification systems are specifically responsible for maintaining xenon purity throughout long-term operation and data-taking. The operational modes are designed to be consistent between the prototype and the future RELICS experiment. A schematic diagram illustrating these modes for both the prototype and the RELICS detectors is shown in Fig.~\ref{fig:relics_prototypes_PID}. 
During the Run~7 prototype operation, a dedicated outgassing vessel was installed between the detector and the purifier to evaluate the efficiency of the purifier. In the standard configuration where this vessel is omitted, the detector is coupled directly to the main circulation loop. This configuration will be employed during the Run~9 prototype operation and future operations. The key components of the entire setup are described as follows:
\begin{itemize}
    
    \item \textbf{Cryogenic System:} 
    This system integrates a Gifford--McMahon (GM) cryocooler (\textit{PengLi KDE400SA} series), an emergency liquid nitrogen (LN$_2$) cooling head and two connected heat exchangers. The GM cryocooler, coupled with a compressor, serves as the primary refrigeration source and maintains the cold-head temperature at approximately \SI{170}{\kelvin} to facilitate xenon liquefaction. During standard operation, purified xenon gas is liquefied at the cold head before entering the TPC. Simultaneously, liquid xenon (LXe) is extracted from the detector and passed through the heat exchange system, where it transfers its cooling capacity to the incoming warm gas, transitioning back into room-temperature gaseous xenon (GXe) for recirculation. In the event of a primary system failure, the LN$_2$ cooling head is activated to provide emergency refrigeration and prevent detector overpressure.

    \item \textbf{Circulation System:} This system is built around a metal bellows pump, mass flow controllers (MFCs), and a high-temperature purifier. The metal bellows pump (\textit{Senior Flexonics MB-602}) ensures a hermetically sealed, oil-free environment, which is critical for preventing the ingress of radon and other atmospheric contaminants. The pump drives the xenon circulation, with the gas flow rate precisely monitored and regulated by \textit{Horiba SEC-N100} MFCs. Within this loop, relatively impure xenon returning from the TPC is driven through the purifier. After the electronegative impurities are removed by the purifier, the clean xenon is returned to the cryocooler or detector, ensuring a continuous purification cycle.

    \item \textbf{Getter-Based Purifier:} The core of the system is a hot zirconium getter (\textit{Simpure 9NG}). As xenon gas flows through the purifier, which is typically operated at 300-400$^\circ$C, electronegative impurities are chemically adsorbed by the getter material. This process is highly effective at removing contaminants like O$_2$ and H$_2$O to below the ppb level.

    \item \textbf{Calibration Sources System:} This system utilizes characteristic energy calibration sources, specifically $^{83\text{m}}\text{Kr}$ and $^{37}\text{Ar}$, which are injected into the circulation loop as internal sources. By mixing these gaseous isotopes with the LXe, the system provides a reliable reference for calibrating the detector response. Analyzing the resulting monoenergetic peaks allows for the accurate determination of the electron lifetime, which is essential for quantifying the electronegative impurity concentrations in the xenon.

\end{itemize}

Building upon the system architecture shown in Fig.~\ref{fig:relics_prototypes_PID}, the xenon circulation modes can consist of several distinct methods of injection into and return from the TPC. They can be selected during operation to meet specific experimental requirements, with their respective flow rates monitored and regulated via MFCs. They are defined as follows, all from the perspective of the TPC:

\begin{itemize}
    \item \textbf{Liquid In (LI):} Purified GXe from the purification system is liquefied at the cold head of the GM cryocooler and subsequently introduced into the liquid xenon at the bottom of the TPC through a dedicated delivery line.

    \item \textbf{Gas In (GI):} Purified GXe from the purification system bypasses the GM cryocooler and is directed into the gas phase at the top of the TPC via a separate injection pathway.

    \item \textbf{Mix In (MI):} Xenon is introduced into the detector through a combination of the LI and GI methods. As these two methods can operate independently from the circulation system, this state represents a mixed injection condition.

    \item \textbf{Liquid Out (LO):} Lower-purity LXe is extracted directly from the TPC and returned to the cryocooler through a dedicated liquid return line, carrying away dissolved impurities. The LXe then enters the heat exchangers, where it vaporizes into the gaseous phase. Ultimately, the vaporized xenon enters the room-temperature circulation loop for purification to remove electronegative impurities.

    \item \textbf{Gas Out (GO):} GXe of relatively lower purity is extracted from the gas phase of the TPC. It is then returned directly to the circulation loop via a separate dedicated gas return line. Once in the circulation loop, GXe passes through the purifier where electronegative impurities are removed.

    \item \textbf{Mix Out (MO):} Xenon is extracted from the detector through both the GO and LO methods simultaneously. Since the two methods operate independently, the xenon flows from both paths eventually merge within the circulation system.
    
\end{itemize}

These injection and return methods can be combined to form complete circulation modes. For example, the Mix In and Mix Out (MI-MO) circulation mode means that xenon is introduced into the detector through mixed injection and return through mixed extraction. Under conditions that maintain stable system operation, various circulation modes can be realized, such as LI-MO and LI-GO.

\subsection{The RELICS Prototype Campaign}
\label{sec:prototype_campaign}

A functionally complete prototype, excluding external shielding, has been constructed for the future RELICS experiment technical validation~\cite{xie2026development}. The prototype integrates the cryogenic, circulation and purification, and calibration subsystems previously described, organized according to the architecture shown in Fig.~\ref{fig:relics_prototypes_PID}. To establish this model, we utilize operational data from two dedicated campaigns, Run~7 and Run~9. During these campaigns, key parameters such as circulation modes and flow rates are modulated to create distinct experimental scenarios, allowing us to assess their respective impacts on purification performance. This subsection details the prototype detector designs for these two runs, as their structural design is closely related to the establishment of the purification model. In addition, it includes the corresponding purification operations and data performance for these two runs, which help us determine the behavior of key parameters.

\subsubsection{Run 7 Configuration and Operation: Diving Bell}

\begin{figure}[htbp]
    \centering
    \includegraphics[width=0.99\linewidth]{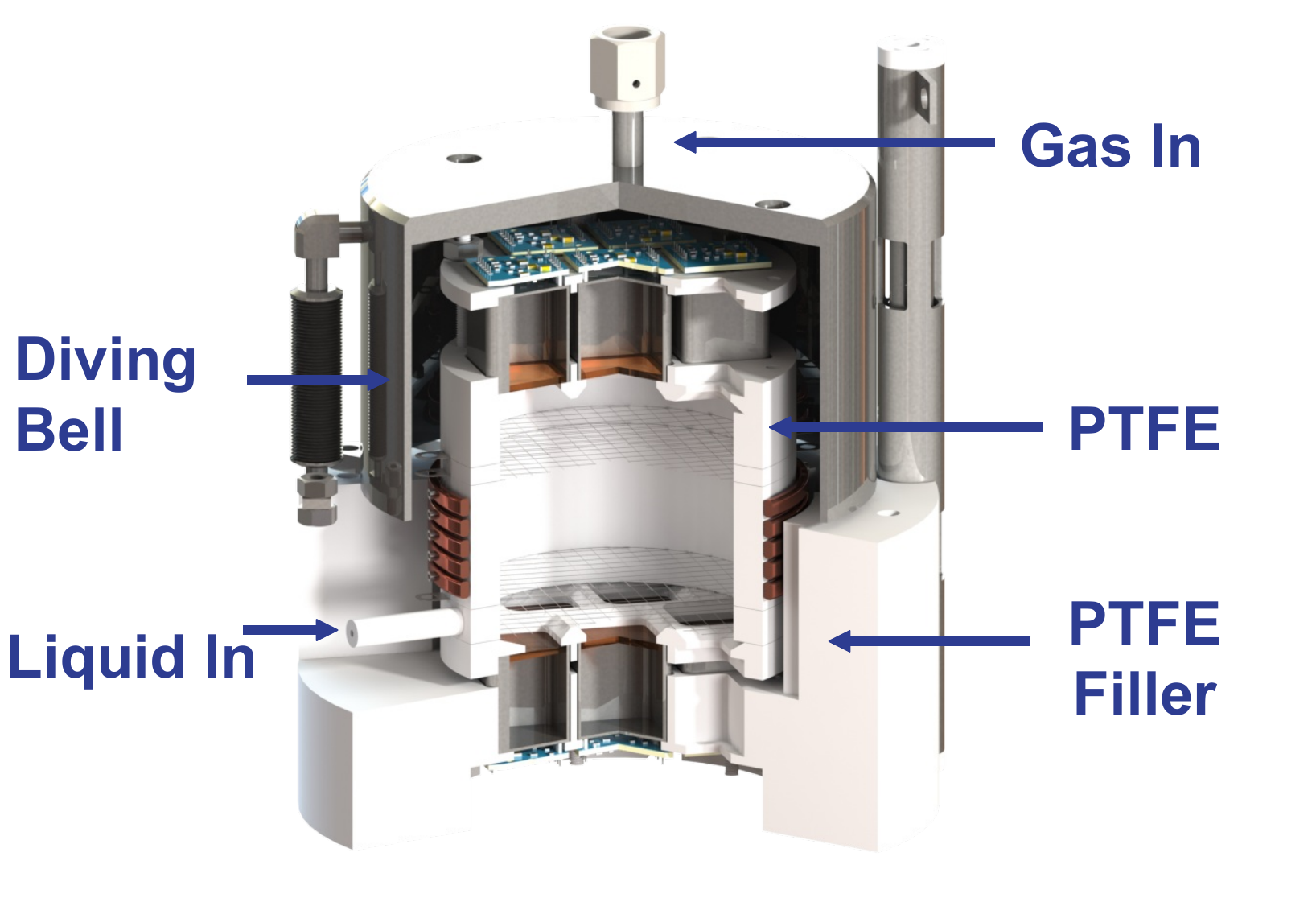}
    \caption{Schematic of the Run~7 prototype configuration. The design features a diving bell structure for liquid level control and includes an internal PTFE filler to reduce the total xenon inventory. Several key structural components and circulation interfaces are indicated in the diagram.}
    \label{fig:tpc_design_run7}
\end{figure}

The Run 7 campaign utilized a TPC designed with a diving bell architecture to precisely control the liquid-gas interface. As detailed in Ref.~\cite{xie2026development} and illustrated in Fig.~\ref{fig:tpc_design_run7}, the TPC features a cylindrical sensitive volume with a diameter of \SI{80}{\milli\meter} and a drift length of \SI{36}{\milli\meter}, containing approximately \SI{0.55}{\kilo\gram} of active LXe. The inner volume is enclosed by high-reflectivity PTFE panels and is instrumented for signal readout with two hexagonal arrays of Hamamatsu R8520-406 Photomultiplier Tubes (PMTs)~\cite{yang2026design}, seven on each of the top and bottom. Surrounding the outer layer of these PTFE panels are copper field-shaping rings, which serve to homogenize the electric field within the drift region.

\begin{figure*}[htbp]
    \centering
    \includegraphics[width=0.99\linewidth]{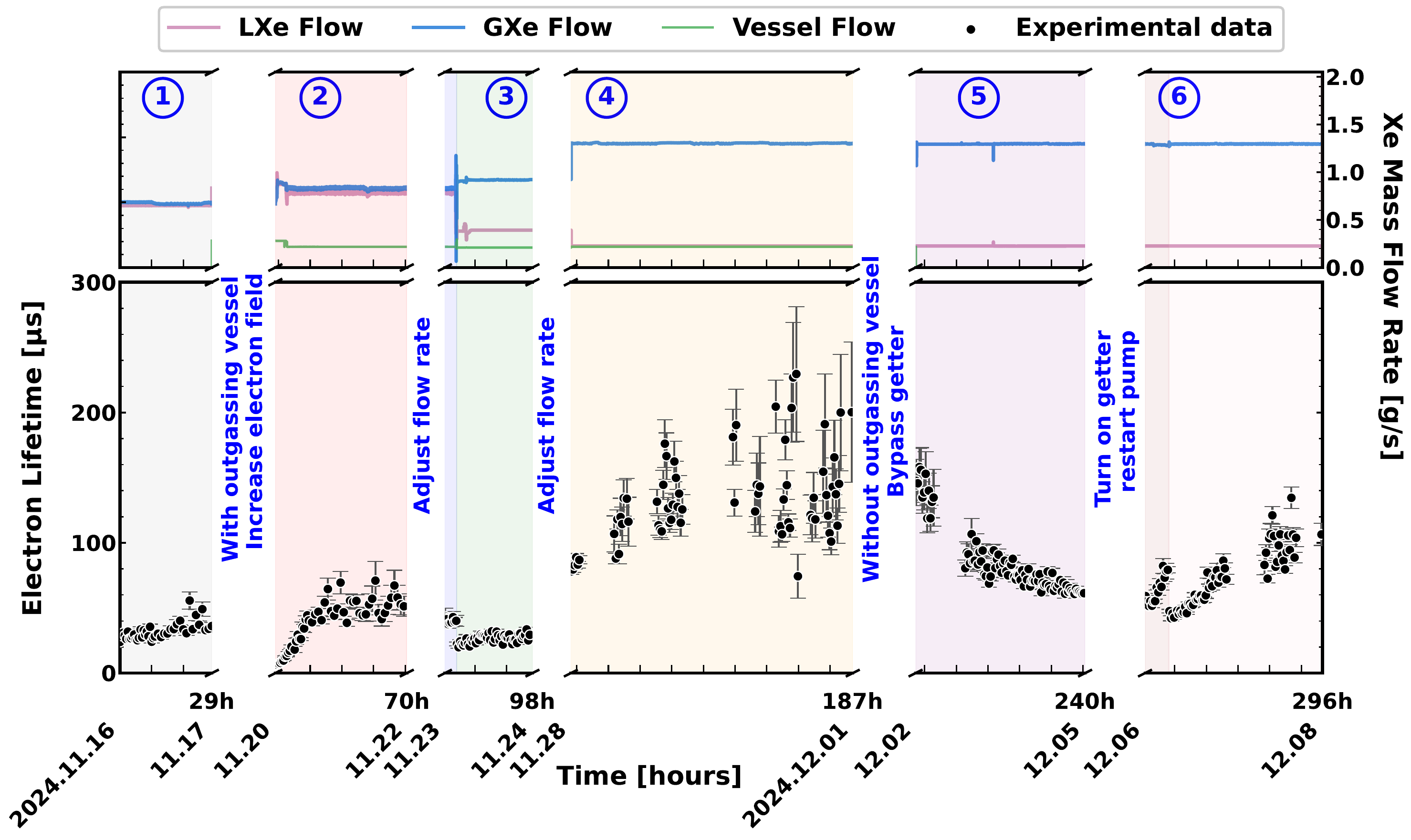}
    \caption{The flow rates during the Run~7 prototype operation and the corresponding electron lifetime data are presented. The upper subfigure shows the xenon mass flow rates in the liquid flow and gas flow modes, indicated by the pink and blue lines, respectively, with units of g/s. The lower subfigure presents the electron lifetime data calculated using the method described in Section~\ref{sec:elife_calculation}, shown as black data points with associated uncertainties, with units of $\mu$s. The blue text annotations highlight key operations, and the colored shaded regions represent different time windows corresponding to different operations during Run~7. }
    \label{fig:operation_run7}
\end{figure*}

A key feature of this configuration is the stainless steel diving bell, which encloses all top PMTs and isolates the gas phase above the TPC sensitive volume from the external gas ullage in the cryostat. A PTFE filler in the annular space outside the diving bell minimizes the total xenon inventory. Purified LXe is delivered into the TPC through a liquid inlet tube and flexible hose, while GXe is supplied to the diving bell via a dedicated gas inlet port. An outlet on the side of the diving bell, leading to the cryostat gas region, vents excess GXe to control the liquid level.

In Run~7, the system employs an MI-MO circulation mode. Separate flow controllers regulate the liquid return branch and the combined gas-liquid delivery line. The diving bell maintains a stable liquid level through active pressure control, implying a balance between gas and liquid inflow and outflow, thereby allowing the circulation rates of both phases to be determined. Notably, a room-temperature gas detector~\cite{guo2026preparation} made of PTFE is installed downstream of the main detector and upstream of the purifier. Within the purification system, this component functions as an outgassing vessel intentionally designed to constrain the purification efficiency of the getter, which represents the getter's capacity to adsorb impurities.

To validate the purification model, a series of controlled operational changes were carried out during Run~7 to vary the circulation conditions and xenon purity. According to different operational and research objectives, the runtime is divided into six phases. The circulation flow rates implemented during these operations and the corresponding purity data are presented in Fig.~\ref{fig:operation_run7}. The horizontal axis of the figure indicates the absolute operation dates, as well as the relative time in hours from the start of the data-taking period, with intervals excluded from the time display.

For Phase~1, the outgassing vessel is disconnected, and both liquid and gaseous xenon are injected and extracted for purification. After Phase~1, the drift electric field is increased from its initial value of \SI{166}{\volt\per\centi\meter} to \SI{333}{\volt\per\centi\meter} and maintained at this higher level for the remainder of the run. Subsequently, to enable verification of the purifier efficiency, an additional outgassing vessel is introduced during Phases~2 to~4. As this vessel is placed upstream of the getter, comparing the purity evolution with and without it constrains the getter's purification efficiency. In addition, to investigate the relationship between purity and circulation flow rate, the liquid and gas circulation flow rates are continuously varied throughout these phases.

In Phase~5, to maintain consistent control variables, the gas and liquid flow rates are kept unchanged. The outgassing vessel is disconnected, and the getter is bypassed. The resulting purity decline, caused by material outgassing within the detector, allows the outgassing rate parameter to be constrained by comparing the observed decrease with independent measurement results. In the final phase, the getter is turned back on, initiating a rise in purity. However, an unexpected stop of the circulation pump interrupts the flow. This interruption may cause liquid level changes in the heat exchanger, introducing impurities into the system and resulting in a purity drop. Following this, the pump is restarted at the original flow rate, and the purity begins to increase again. Moreover, in Phase~6, where no outgassing vessel is connected, further comparison with the previous data can be performed to constrain the purification efficiency.

The evolution of purity resembles a Fermi-Dirac distribution. During the purification phase in which impurities are adsorbed by the getter, the purity increases, exhibiting a rapid initial rise followed by a slow increase before eventually entering a plateau phase. This rising phase is mainly influenced by the purification efficiency and flow rate. When the impurity reduction rate due to purification approaches the impurity increase rate due to material outgassing, the purity ceases to increase. At this point, if purification is stopped, the material outgassing rate leads to a decline phase characterized by a rapid initial drop followed by a slow decrease. Our continuous changes to the detector state are intended to obtain these data features in order to better establish the purification model.

\subsubsection{Run 9 Configuration and Operation: Overflow Chamber}

\begin{figure}[htpb]
    \centering
    \includegraphics[width=0.99\linewidth]{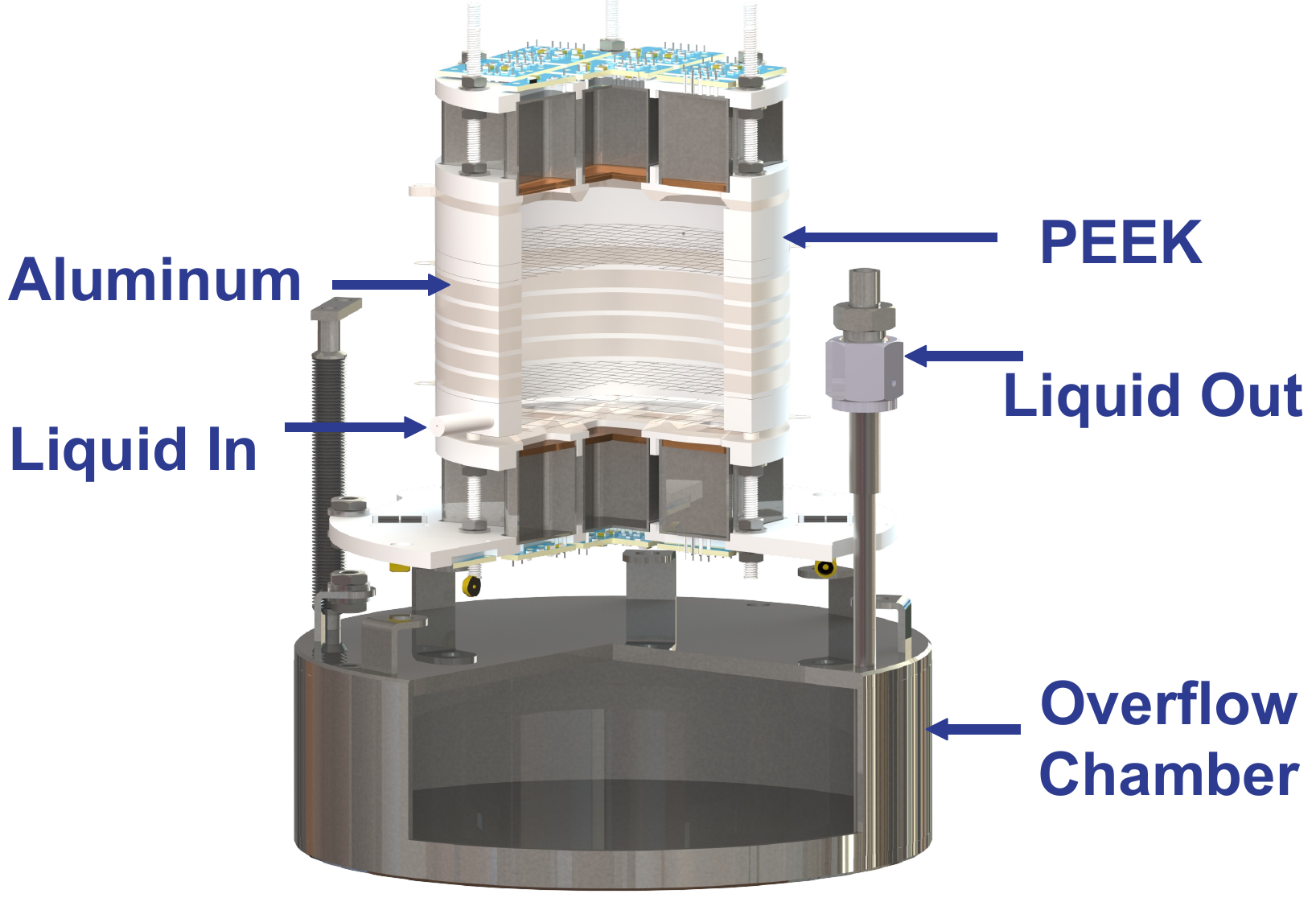}
    \caption{Schematic of the Run~9 prototype configuration. The design utilizes an overflow chamber for passive liquid level control, and the key TPC components are constructed from aluminum and PEEK to serve as reflective materials. Several important structural features and interfaces are also indicated.}
    \label{fig:design_tpc_run9}
\end{figure}

The Run~9 prototype employs an overflow chamber design instead of the diving bell configuration used in Run~7, while also serving to further validate the purification model. The Run~9 campaign maintains major performance parameters, including the number of PMTs, the drift region length and radius, and the resulting sensitive volume, consistent with the Run~7 configuration. In addition, the choice of reflective material differs between the two runs. As illustrated in Fig.~\ref{fig:design_tpc_run9}, the internal TPC components are fabricated primarily from polyether ether ketone (PEEK) and aluminum (Al). Distinct from the Run~7 design, spaced Al strips are employed as field-shaping rings to homogenize the electric field, and they are in direct contact with the LXe volume.

\begin{figure*}[htbp]
    \centering
    \includegraphics[width=1\linewidth]{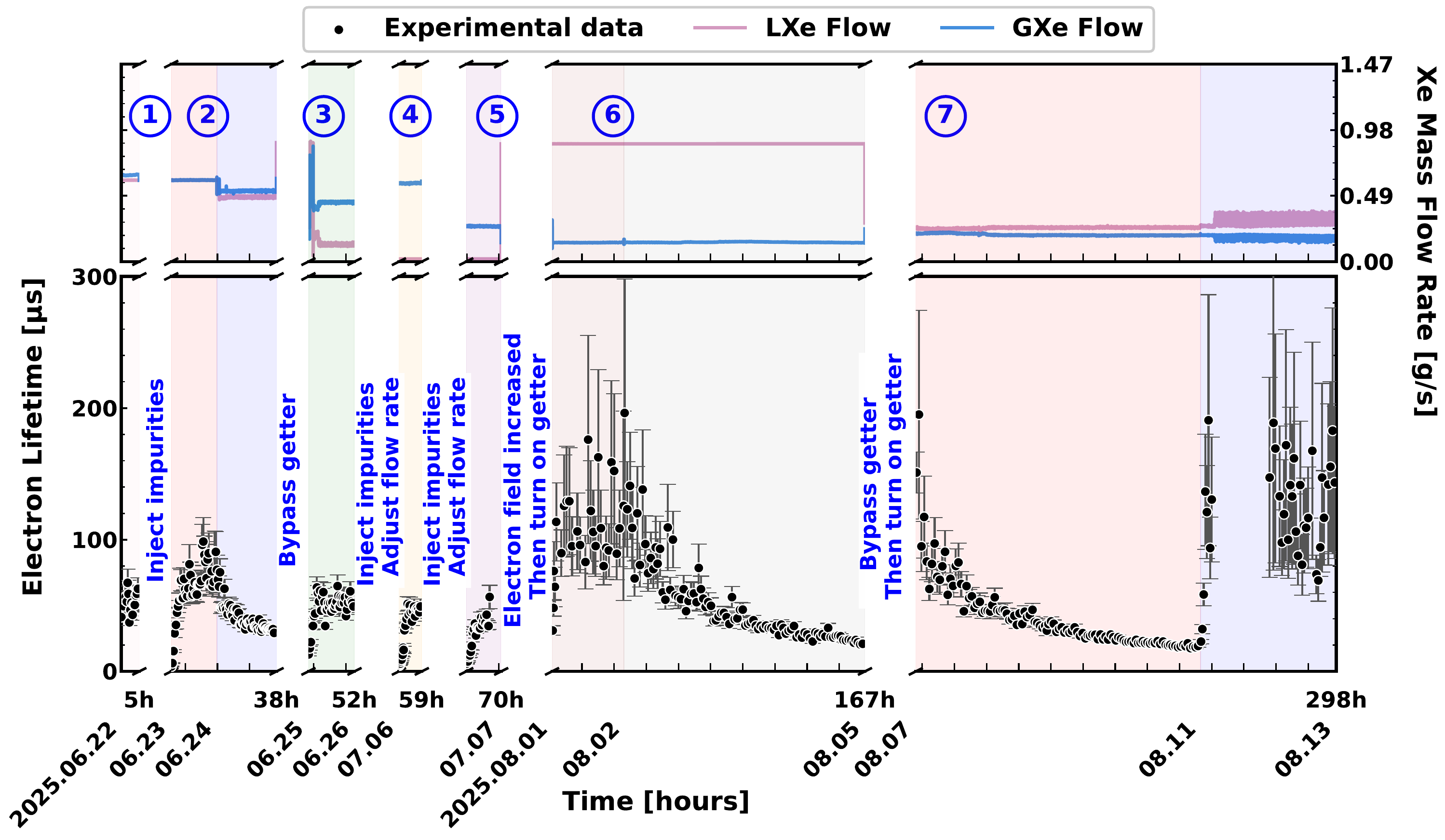}
    \caption{The flow rates during the Run~9 prototype operation and the corresponding electron lifetime data are presented. The upper subfigure shows the xenon mass flow rates in the liquid flow and gas flow modes, indicated by the pink and blue lines, respectively, with units of g/s. The lower subfigure presents the electron lifetime data calculated using the method described in Section~\ref{sec:elife_calculation}, shown as black data points with associated uncertainties, with units of $\mu$s. The blue text annotations highlight key operations, and the colored shaded regions represent different time windows corresponding to different operations during Run~9. }
    \label{fig:operation_run9}
\end{figure*}

In this configuration, liquid level is passively regulated by the overflow chamber. LXe is delivered into the sensitive volume via the liquid inlet port shown in Fig.~\ref{fig:design_tpc_run9}. In practice, the airtight connection between flexible hoses at this port is reinforced using VCR fittings. Excess liquid spills over a fixed weir into a dedicated reservoir before being extracted via the liquid return port indicated in the figure. This standalone stainless steel overflow chamber introduces its own outgassing rate, thereby affecting the purity. Gas is extracted from the thermostatic tank and returned to the circulation loop through the gas return line. This passive overflow mechanism maintains a constant liquid level without active control, effectively decoupling level regulation from circulation and significantly improving operational stability.

In the Run~9 operation, the system operates in the LI-MO circulation mode. Since the liquid level is regulated by the overflow chamber, active pressure control within the detector is no longer required to maintain the liquid level; consequently, the GI inlet is omitted. Because the overflow chamber offers greater operational flexibility, we purposefully vary the operating conditions to obtain different purity characteristics, and the runtime is divided into seven phases. Figure~\ref{fig:operation_run9} shows the circulation flow rates used for these operations and the resulting purity data. A detailed summary of the operations and timeline during Run~9 is as follows.

For Phase~1, the purification begins after xenon filling with the getter on, and during this first circulation and purification stage, the purity rises slowly due to residual impurities present in some parts of the system. Then, in order to obtain a more complete purification rising phase, a controlled amount of impurities is injected in Phase~2 after bypassing the getter, thereby causing the purity to drop rapidly to a low level. Subsequently, following the reconnection of the getter, the purity rises and reaches a stable plateau, and after stabilization, the getter is bypassed again, thus allowing the purity of the LXe to decline due to material outgassing. In Phases~3,~4, and~5, the above procedure of impurity injection, purity reduction, flow rate adjustment, and re-purification to achieve a purity rising phase is repeated.

For Phases~5 and~6, the drift electric field is increased from \SI{200}{\volt\per\centi\meter} to \SI{500}{\volt\per\centi\meter}. A higher drift field leads to a greater electron drift velocity and a smaller attachment rate, while also mitigating ionization yield variations caused by electric field non-uniformity. Data acquired under this modified field condition are compared with those from the prior, lower-field operation. Apart from the field-dependent variables, parameters governing purification efficiency and outgassing rates during operation are expected to remain unchanged. In Phase~5, the circulation flow rate is kept constant: the purity first rises through purification, and then the getter is bypassed, causing the purity to decline. After the purity reaches the plateau phase, the circulation flow rate is adjusted to a lower value in Phase~6, and the getter is bypassed again to obtain another purity decline phase. After the purity slowly declines to a low level, the getter is reconnected to the system to obtain a continuous purity ascent phase. Under the lower gas and liquid return flow rates, some flow fluctuations occur due to the circulation pump, but the system state remains relatively stable.

The basic approach follows the same logic as before. This run includes direct comparisons under different circulation flow rates and introduces more continuous changes. This sequence of continuous ascent and descent data is advantageous for determining the purification efficiency and the material outgassing rate, as it involves fewer initial concentration parameters and thus reduces the number of variables.

Consequently, a complete set of purification operational data, measured by the electron lifetime, is obtained from both the Run~7 and Run~9 prototype operations, including adjustments to the circulation flow rate, connecting or bypassing the getter, and other operational variations. Since the prototype tests also serve other physical validation purposes, certain operational adjustments inevitably impact LXe purity or the measurement of electron lifetime. The purification model will subsequently be used to describe the characterization of these data and the effects of each operation.

\subsection{Acquisition of Prototype Purity Data}
\label{sec:elife_calculation}

Building upon an understanding of the prototype's structure and operational modes, the purity of LXe is derived through the analysis of the ionization signals recorded by the prototype detector. Two types of signals are recorded in the TPC: the primary scintillation signal (S1) and the amplified ionization signal (S2) generated via electroluminescence. The ionization signal originates from the electrons created by the energy deposition at the vertex of particle interaction. These electrons drift under an applied electric field towards the liquid-gas interface, where they are extracted and amplified in the GXe. During this drift, electronegative impurities present in the LXe can capture electrons, leading to an attenuation of the signal. The characteristic scale of this attenuation, quantified by the electron lifetime $\tau_e$, serves as the direct measure of the LXe purity. Thus, $\tau_e$ is inversely proportional to the impurity concentration, governed by the relation:

\begin{equation}
{\tau_e} = \frac{1}{k \cdot I_{[\text{O}_2]}},
\end{equation}
where $k$ is the electron attachment rate coefficient with units of \si{\liter\per\mole\per\second}, and $I_{{[\text{O}_2]}}$ is the concentration of electronegative impurities with units of \si{\mole\per\liter} \cite{elife_02}. It is important to note that $k$ itself is inversely dependent on the electric field strength. A stronger field reduces $k$ because it increases the electron drift velocity, thereby reducing the effective interaction time with impurities and lessening signal loss.

The attenuation of the electron signal due to impurity capture follows an exponential decay as a function of drift time:
\begin{equation}
    N(t_z) = N_0 \cdot e^{-t_z / \tau_e},
\end{equation}
where the electron drift time $t_z$ is determined from the time difference between the prompt S1 and the delayed S2. Since the amplitude of the S2 signal is proportional to the number of electrons $N$ that reach the liquid-gas interface, the decay can be expressed directly using the measured signals, with $N(t_z)$ and $N_0$ represented by the observed $S2(t_z)$ and the extrapolated initial amplitude $S2_0$ (at $t_z = 0$), respectively. Therefore, the above formula can be rewritten as:
\begin{equation}
\label{eq:elife_s2}
    \lg S2(t_z) = \lg S2_0 - \frac{1}{\tau_e} t_z.
\end{equation}

\begin{figure}[htpb]
    \centering
    \includegraphics[width=0.99\linewidth]{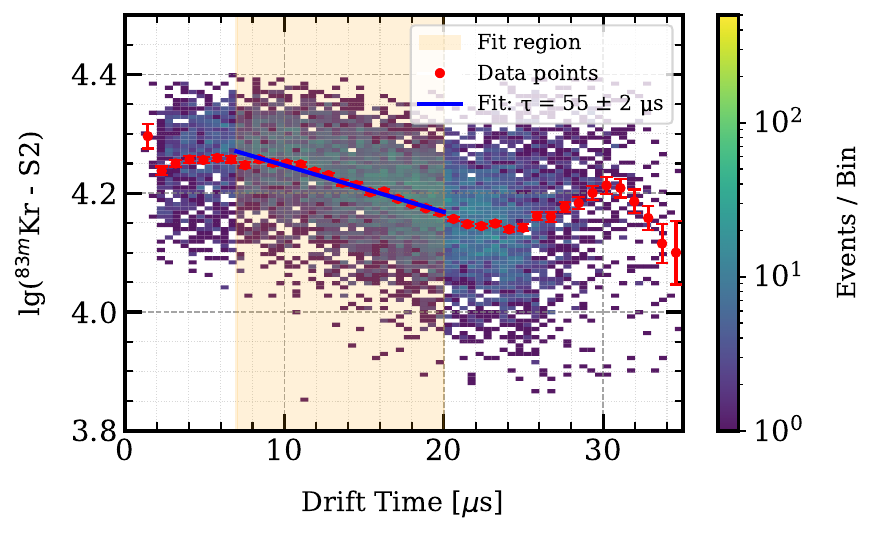}
    \caption{The electron lifetime $\tau_e$ is extracted by fitting the S2 signal amplitude as a function of drift time for events selected within the monoenergetic $^{83\text{m}}$Kr band, where the yellow-shaded region marks the drift times corresponding to the central selection of the detector, and the red points represent the mean S2 amplitude for each drift time bin.}
    \label{fig:elife_calculation_k}
\end{figure}

So $\tau_e$ is obtained from the negative reciprocal of the slope in a linear fit to the dependence of $\lg(S2)$ on the electron drift time $t_z$, as shown in Fig.~\ref{fig:elife_calculation_k}. By repeating this procedure for data acquired at different times, the temporal evolution of the electron lifetime is obtained. Since this method relies on data acquired within the TPC, the purity derived from $\tau_e$ is inherently influenced by the uniformity of the electric field in the active volume. Consequently, electric field non-uniformity can introduce systematic effects into the measurement of $\tau_e$. This factor should be accounted for when comparing results from different detector configurations (e.g., Run~7 and Run~9).

\begin{figure}[htpb]
    \centering
    \includegraphics[width=0.99\linewidth]{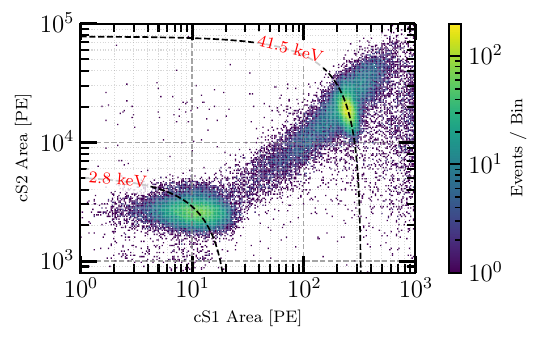}
    \caption{Two-dimensional distribution of the S2 versus S1 signal areas from events in the prototype detector. The clusters correspond to monoenergetic events from the $^{83\text{m}}$Kr and $^{37}$Ar calibration sources, respectively~\cite{xie2026development}.}
    \label{fig:elife_calculation}
\end{figure}

In the prototype data analysis, both the 2.8~keV line from $^{37}$Ar and the 41.5~keV line from $^{83\mathrm{m}}$Kr are used for energy calibration, and they can also be used for electron lifetime calculation. In this work, $^{83\text{m}}$Kr events are selected as a calibration source to determine the electron lifetime ~\cite{zhang2022rb,be2013table}. The $^{83}$Rb isotope, which has a relatively long half-life of about 86 days, decays to produce $^{83\text{m}}$Kr. This decay chain maintained a sufficiently high activity throughout both prototype Run~7 and Run~9, providing ample statistical precision for the calculation. The $^{83\text{m}}$Kr decay provides a well-defined energy line at 41.5~keV, ensuring that the central value of $S2_0$ is consistent across events and that its amplitude variation is influenced primarily by statistical fluctuations and electron attenuation during drift. In fitting the electron lifetime, taking the mean S2 amplitude within bins of drift time mitigates statistical fluctuations in the energy-to-signal conversion.

To quantitatively describe the relationship between the electron lifetime and the detection threshold, we introduce a variable defined as the electron survival probability \( P_{z} \). Following the description above, it can be expressed as:

\begin{align*}
P_{z} = e^{-t/\tau} = e^{-z / (v \cdot \tau)}
\end{align*},
where \( t \) is the electron drift time, \( \tau \) is the electron lifetime, \( z \) is the drift distance, and \( v \) is the drift velocity. This formulation allows us to directly evaluate the fraction of ionization electrons that survive to the liquid-gas interface and contribute to the S2 signal. For low-threshold signal detection, high purity and long electron lifetime are essential to ensure that low-energy events with few electrons can drift to the amplification region, thereby maintaining a low detection threshold. Meanwhile, events occurring near the bottom of the detector, which have longer drift times, can also retain high signal integrity, ensuring a sufficiently large sensitive volume.

\section{Development and Validation of Purification Model }
\label{sec:purity}

To determine the future circulation and purification system design for the RELICS experiment, such as appropriate circulation flow rates, it is necessary to understand the evolution of liquid xenon purity and quantify the impact of various parameters on purity. This section first describes the prototype detector design and operational conditions, based on which a comprehensive purity evolution model has been developed. Subsequently, by utilizing the evolution of purity level during the Run~7 and Run~9 operations of the prototype detector, this purification model is validated. This model serves to guide the design of the future RELICS detectors, along with their associated purification and circulation systems. In addition, we present the results of material outgassing measurements, which provide the foundational inputs for the model. Finally, based on this model, predictions for the purity performance of the future RELICS experiments are presented.

\subsection{Purification model}
\label{sec:model_all}
To quantitatively model the temporal evolution of purity within the detector, a comprehensive purification model has been developed. This model is designed to account for all relevant physical parameters and their respective influences. Given the non-uniform conditions present in the system, such as varying temperatures, phases, and flow rates, the detector is conceptually divided into multiple interconnected and well-mixed volumes. Xenon, along with impurities, exchanges between these volumes through continuous processes including liquefaction, evaporation, and circulation, maintaining a dynamic equilibrium. This framework allows for a detailed description of impurity transport and distribution. The model integrates the full detector design, as detailed in Section~\ref{sec:prototype_campaign}, with the specific operations and real-time system status. By coupling these elements, the model provides a time and parameter-dependent simulation of impurity concentration evolution throughout the entire cryogenic and purification system.

\subsubsection{Purification Model for Run 7}
\label{sec:run7_model_quaction}

In Run~7, impurity exchange occurs between these volumes as circulation proceeds, allowing for the formulation of evolution equations that describe the impurity concentration in each region. Based on the actual detector design, which features the diving bell structure as shown in Fig.~\ref{fig:tpc_design_run7}, six distinct regions are defined, with xenon and impurity exchange occurring between them.

\begin{figure}[htbp]
    \centering
    \includegraphics[width=0.99\linewidth]{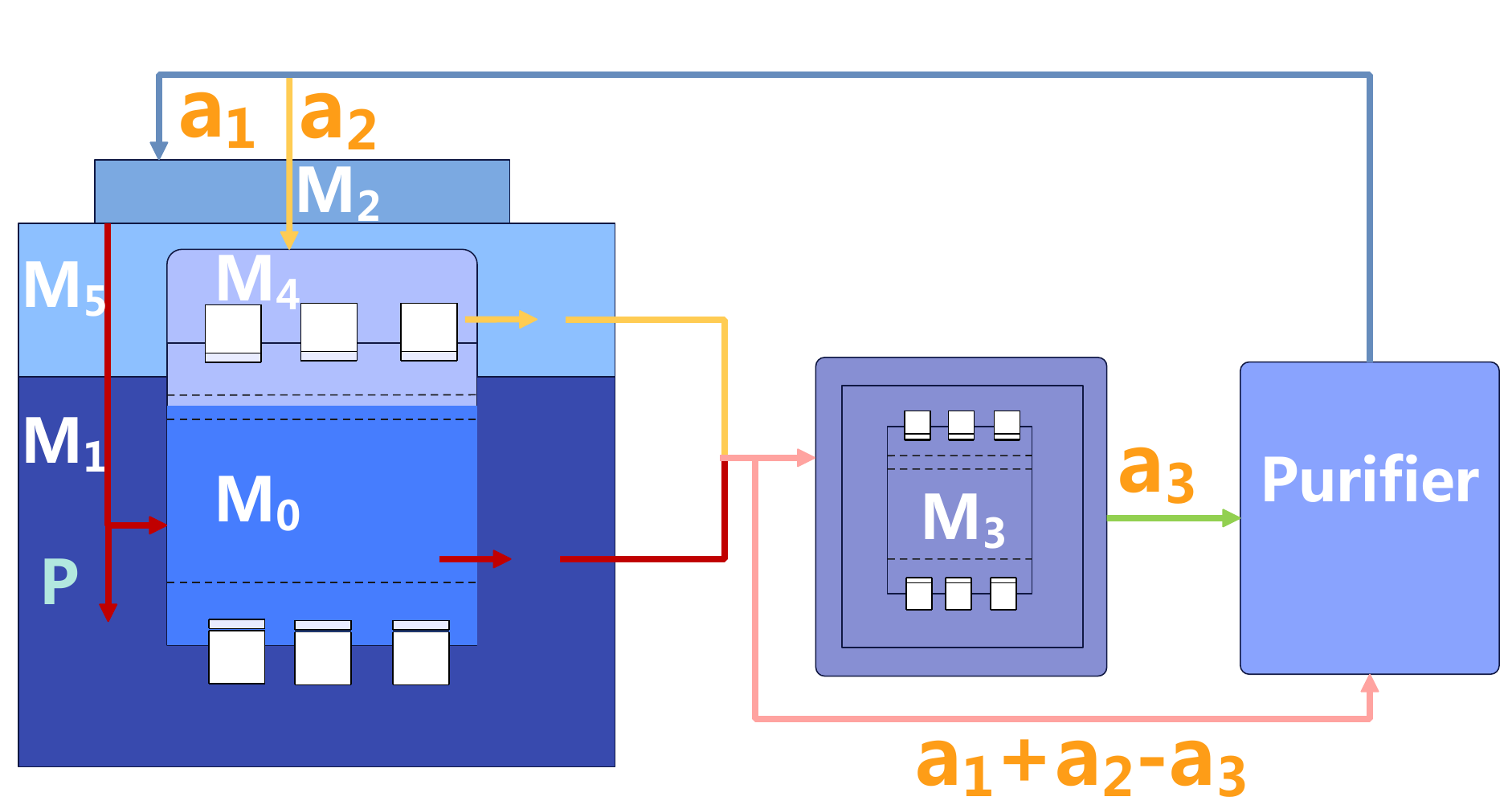}
    \caption{Schematic diagram of the various regions in the detector, where $M$ denotes the mass of different regions, $a$ is the circulation flow rate, and the arrows indicate the injection and extraction of xenon and exchange between different regions.}
    \label{fig:model_run7}
\end{figure}

The impurity evolution in each region is described by the following differential equations \eqref{eq:impurity_evolution_run7}, which model the impurity concentration dynamics and circulation processes for the regions defined in Fig.~\ref{fig:model_run7}.

\begin{equation}
\label{eq:impurity_evolution_run7}
    \left\{
    \begin{aligned}
    & \begin{multlined}[t]
      \text{LXe within TPC - } M_{0}: \\[8pt]
      \quad n_0 \frac{dx_0}{dt} = \frac{n_0}{M_0}\left[ {p  a_1} x_{\mathrm{2}} - {p  a_1} x_0 \right] + f_0 - T_0
      \end{multlined} 
    & \\[6pt]
    & \begin{multlined}[t]
      \text{LXe out TPC - }M_{1}: \\[8pt]
      \quad n_1 \frac{dx_1}{dt} = \frac{n_1}{M_1} \left[ {p  a_1}x_0 + {(1-p)  a_1}x_{\mathrm{2}} - {a_1} x_1 \right] \\ 
      + f_1 - T_1
      \end{multlined} 
    & \\[6pt]
    & \begin{multlined}[t]
      \text{GXe within GM cryocooler - }M_{2}: \\[8pt]
      \quad n_2 \frac{dx_2}{dt} = \frac{n_2}{M_2} \left[ {a_1} x_{\text{mix}} - {a_1} x_2 \right] + f_2
      \end{multlined} 
    & \\[6pt]
    & \begin{multlined}[t]
      \text{GXe within outgassing vessel - }M_{3}: \\[8pt]
      \quad n_3 \frac{dx_3}{dt} = \frac{n_3}{M_3} \left[{a_3} x_{\text{mix-c}} - {a_3}x_3 \right] + f_3
      \end{multlined} 
    & \\[6pt]
    & \begin{multlined}[t]
      \text{GXe within TPC - } M_{4}: \\[8pt]
      \quad n_4 \frac{dx_4}{dt} = \frac{n_4}{M_4}\left[ {a_2}x_{\mathrm{mix}} - {a_2}x_4 \right] + f_4 + T_0
      \end{multlined} 
    & \\[6pt]
    & \begin{multlined}[t]
      \text{GXe out TPC - } M_{5}: \\[8pt]
      \quad n_5 \frac{dx_5}{dt} = \frac{n_5}{M_5} \left[ {a_2} x_4 - {a_2} x_5 \right] + f_5 + T_1
      \end{multlined}
    \end{aligned}
    \right.
\end{equation}

The entire circulation process can be described as follows. Purified gaseous xenon splits into two parallel flows. The LI flow branch directs gas to the GM cryocooler volume ($M_2$), where it is liquefied. The resulting LXe is then transferred via the liquid delivery line into the TPC's internal LXe region ($M_0$). To maintain liquid level stability, excess xenon in $M_0$ overflows into the external liquid reservoir ($M_1$). Additionally, due to potential sealing inefficiencies, a portion of the liquid may bypass the TPC and enter $M_1$ directly. The LO line, which is equipped with a dedicated flow controller, extracts xenon from $M_1$ at a measured flow rate $a_1$. After vaporization in the cryocooler's heat exchanger, this flow returns to the circulation system. Concurrently, the GI branch supplies gas directly into the diving bell ($M_4$) to replenish the gaseous xenon volume within the TPC. Because the pressure in $M_4$ regulates the liquid level position, any excess gas is vented into the external gaseous region ($M_5$) via an exhaust line. The gas return line then extracts xenon from $M_5$ and returns it to the circulation loop.

The xenon flows from both GO and LO converge and are thoroughly mixed. It is important to note that prior to entering the purifier, a portion of this mixed flow may be diverted through a dedicated outgassing vessel ($M_3$) for the specific purpose of validating the getter's purification efficiency. This line is also equipped with a flow controller, providing the flow rate $a_3$ through the outgassing vessel. The remainder of the flow proceeds directly to the getter. Both flows are ultimately mixed inside the getter, where electronegative impurities are removed. The purified gas then re-enters the cycle at the split point, completing the closed-loop circulation. The total circulation flow rate $a_{\text{total}}$ is regulated by another master flow controller. The gas circulation flow rate $a_2$ is subsequently determined by subtracting the measured liquid return flow rate $a_1$ from this total: $a_2 = a_{\text{total}} - a_1$.

The symbols used in the following equations are defined as follows. For each volume indexed by $i$, $M_i$ (in kg) denotes the xenon mass contained within that volume, and $a_i$ (in kg/s) represents the associated circulation flow rate entering or leaving it. The molar amount of xenon is denoted by $n_i$ (in mol), and $x_i$ is the dimensionless impurity-to-xenon ratio in volume $i$. The parameter $f_i$ (in ppb mol/h) denotes the material outgassing rate into volume $i$, which is determined by the material type, surface area, and temperature specific to that region. In a closed and air-free system, the outgassing rate for a given material typically decays over time. However, for the purpose of this model and to maintain a steady-state baseline for the purity evolution, the value of $f_i$ is treated as a constant. This constant value is chosen to represent the outgassing rate at the beginning of xenon filling. 

For consistent unit handling, both sides of the equation are expressed in mol/s. Finally, the impurity concentration \(I_i\) in each volume is given by:
\begin{equation}
I_i = \frac{n_i \cdot x_i}{V_i},
\end{equation}
where \(V_i\) (in L) represents the volume of each region.

Additionally, $p$ is a dimensionless parameter ($0 \leq p \leq 1$) describing the efficiency of the liquid delivery line entering the TPC. Ideally, to maximize the purity within $M_0$, a value of $p$ close to 1 is desired, meaning all incoming purified liquid enters $M_0$ directly. However, this efficiency depends on the actual detector design and sealing, and any inefficiency ($1-p$) results in a bypass flow to $M_1$. This effect is accounted for in the mass balance equations for volumes $M_0$ and $M_1$.

Furthermore, the terms $T_0$ and $T_1$ in the equations represent the gas-liquid impurity exchange at the liquid-gas interface. According to Henry's law, impurities tend to migrate from regions of higher to lower concentration. Due to the much faster circulation in the gaseous phase compared to the liquid phase, the impurity concentration in the gas is typically lower. Therefore, wherever a stable gas-liquid interface exists, an impurity exchange flux across that interface must be considered. In our detector design, the diving bell structure isolates $M_4$ from $M_5$. Consequently, we only need to consider two distinct gas-liquid exchange terms: $T_0$ across the interface between  $M_0$ and $M_4$, and $T_1$ across the interface between  $M_1$ and $M_5$. These exchange terms are modeled as:
\begin{equation}
\begin{aligned}
T_{0} &=  h \cdot A_{\mathrm{TPC}} \cdot \left(K_H \cdot \frac{\rho_g}{M_{\mathrm{Xe}}} \cdot x_0 - I_{4} \right) \\[10pt]
T_{1} &=  h \cdot A_{\mathrm{Out}} \cdot \left(K_H \cdot \frac{\rho_g}{M_{\mathrm{Xe}}} \cdot x_1 - I_{5} \right),
\end{aligned}
\end{equation}
where $A_{\mathrm{TPC}}$ and $A_{\mathrm{Out}}$ (in $\text{m}^2$) are the surface areas of the liquid in the TPC and the external reservoir, respectively, calculated from the detector geometry. The parameter $h$ is the convective mass transfer coefficient (in $\text{m/s}$) ~\cite{Plante:2022khm}. This coefficient is related to heat transfer, and differences in heat transfer between liquid xenon regions caused by the detector structure affect the strength of gas-liquid exchange. Henry's law volatility constant for O$_2$ in xenon is $K_H = 62.5$~\cite{o2henry}. The gaseous xenon density is $\rho_g = 5.89\ \text{kg/m}^3$, and $M_{\mathrm{Xe}} = 131\ \text{g/mol}$ is the molar mass of xenon.

The $M_2$ region is where gaseous xenon undergoes liquefaction and cold head of the cryocooler is located in this volume. Under certain conditions, impurities can freeze onto the surface of the cold head. This phenomenon introduces an additional impurity source beyond the constant baseline material outgassing rate ($f_2$), which persists until the frozen impurities eventually sublimate or melt. We define this additional impurity source term as $f_{\mathrm{coldhead}}$. Therefore, during periods when ice builds up on the cold head, the impurity mass balance equation for volume $M_2$ must be modified to account for this extra source. The governing equation becomes:
\begin{equation}
n_2 \frac{dx_2}{dt} = \frac{n_2}{M_2} \left[ {a_1} x_{\text{mix}} - {a_1} x_2 \right] + f_2 + f_{\mathrm{coldhead}}.
\end{equation}

Within the dynamic equilibrium of the circulation loop, xenon returning from both the liquid and gas phases must be thoroughly mixed before re-entering the purification stage. This complete mixing of the return flows is represented in the model by the impurity concentration $x_{\mathrm{mix-c}}$, calculated as the flow-weighted average:
\begin{equation}
x_{\mathrm{mix-c}} = \frac{a_1 \cdot x_1 + a_2 \cdot x_5}{a_1 + a_2}.
\end{equation}

Subsequently, a fraction of this mixed flow, at a rate $a_3$, is diverted through the $M_3$, while the remainder proceeds directly to the getter. Crucially, both flows converge and are completely mixed within the getter itself. The final impurity concentration $x_{\mathrm{mix}}$ of the gas exiting the getter and ready to be re-injected into the system accounts for this merging and the purification efficiency $e$ of the getter:
\begin{equation}
x_{\mathrm{mix}} = \left[ \frac{(a_1 + a_2 - a_3) \cdot x_{\mathrm{mix-c}} + a_3 \cdot x_3}{a_1 + a_2} \right] (1 - e),
\end{equation}
where, $e$ (with $0 \leq e \leq 1$) denotes the purification efficiency of the getter, representing the fraction of impurities removed. Consequently, the fraction $(1-e)$ of impurities that are not removed re-enters the circulation system with the purified gas.

Another factor that needs to be considered is the explicit inclusion of a parameter to account for the impact of electric field distortion. In practice, as mentioned earlier, the LXe purity is inferred from the measured electron lifetime $\tau_m$, the calculation of which depends on the signal amplitude, specifically the electron yield. Electric field non-uniformity, which typically manifests as a stronger field at the cathode (bottom) and a weaker field at the anode (top) of the drift region, introduces a systematic upper limit to the measurable electron lifetime. This upper limit is intrinsically linked to the design of the field-shaping rings and the overall detector structure. To incorporate the impact of this electric field non-uniformity into the model, the following modified equation is introduced:
\begin{equation}
\frac{1}{\tau_{\mathrm{m}}} = \frac{1}{\tau_{\mathrm{true}}} + \frac{1}{\tau_E},
\end{equation}
where $\tau_{\mathrm{m}}$ is the measured electron lifetime, $\tau_{\mathrm{true}}$ is the lifetime limited solely by impurity concentration, and $\tau_E$ represents the effective lifetime reduction induced by field non-uniformity. The term $\tau_E$ is related to the overall charge collection efficiency under the specific non-uniform field configuration and therefore depends on the detector geometry and applied voltage. The physical interpretation of this equation is clear: In the case of a perfectly uniform electric field, $\tau_E \to \infty$, and thus $\tau_{\mathrm{m}} \approx \tau_{\mathrm{true}}$, meaning the measurement is unaffected by field distortions. Conversely, under significant field non-uniformity, $\tau_E$ becomes a finite, dominant term, imposing a strict upper limit on the measurable lifetime such that $\tau_{\mathrm{m}} \le \tau_E$.

Another notable point is that certain special cases occur during operation. For example, in Run~7, the circulation pump stopped unexpectedly. We speculate that this caused a change in the liquid level within the cryocooler heat exchanger, thereby introducing some impurities. Other cases include intentional impurity injection to achieve low concentration levels, as well as other discontinuous operations. In all such cases, the purification model introduces new $x_i$ parameters as new starting points, while keeping other parameters related to detector performance unchanged throughout the same Run.

\subsubsection{Purification Model for Run 9}
\label{sec:run9_model_equation}

In the Run~9 configuration, the diving bell is replaced with an overflow chamber, as shown in Fig.~\ref{fig:design_tpc_run9}. Consequently, the system is partitioned into five control volumes, and the room-temperature gas detector is omitted. While the circulation system maintains the circulation mode as LIMO or LIGO and the overall process and volume definitions remain largely consistent with Run~7, the mechanism for impurity exchange is revised due to structural changes.

\begin{figure}[htbp]
    \centering
    \includegraphics[width=0.99\linewidth]{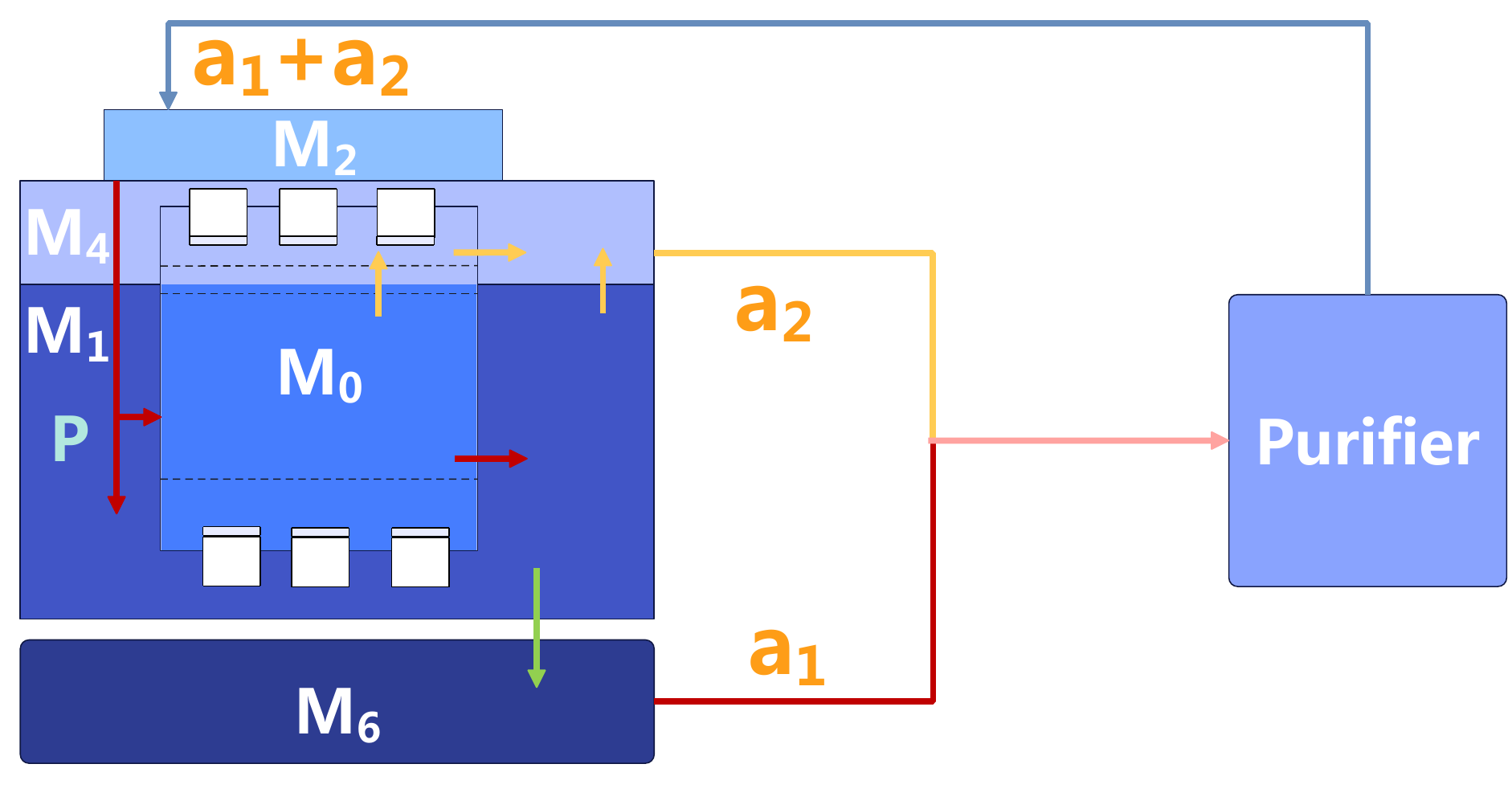}
    \caption{Schematic diagram of the various regions in the detector, where $M$ denotes the mass of different regions, $a$ is the circulation flow rate, and the arrows indicate the xenon sources and exchange between different regions.}
    \label{fig:model_run9}
\end{figure}

The following differential Eq.~\eqref{eq:impurity_evolution_run9} models the impurity concentration dynamics and circulation processes for the regions illustrated in Fig.~\ref{fig:model_run9}.

\begin{equation}
\label{eq:impurity_evolution_run9}
    \left\{
    \begin{aligned}
    & \begin{multlined}[t]
      \text{LXe within TPC - } M_{0}:\\[8pt]
      \quad n_0 \frac{dx_0}{dt} = \frac{n_0}{M_0} \bigl[ p(a_1+a_2)(x_{\mathrm{2}} - x_{\mathrm{0}} )] + f_0 - T_0
      \end{multlined} 
    &\\[6pt]
    & \begin{multlined}[t]
      \text{LXe out TPC - } M_{1}: \\[8pt]
      \quad n_1 \frac{dx_1}{dt} = \frac{n_1}{M_1}\bigl[ p(a_1+a_2)x_0 + (1-p)(a_1+a_2)x_{\mathrm{2}} \\[8pt] - (a_1+a_2)x_1 \bigr] + f_1 - T_1
      \end{multlined} 
    & \\[6pt]
    & \begin{multlined}[t]
      \text{GXe within GM cryocooler - } M_{2}: \\[8pt]
      \quad n_2 \frac{dx_2}{dt} = \frac{n_2}{M_2} \bigl[ (a_1+a_2)x_{\text{mix}} - (a_1+a_2)x_2 \bigr] + f_2
      \end{multlined} 
    & \\[6pt]
    & \begin{multlined}[t]
      \text{GXe within tank - } M_{4}: \\[8pt]
      \quad n_4 \frac{dx_4}{dt} = \frac{n_4}{M_4}\Bigl[ \frac{a_2}{A_{\text{in}}+A_{\text{out}}} (A_{\text{in}}x_0 + A_{\text{out}}x_1) - a_2x_4 \Bigr] \\[8pt] + f_4 + T_0 + T_1
      \end{multlined} 
    & \\[6pt]
    & \begin{multlined}[t]
      \text{LXe in overflow chamber - } M_{6}: \\[8pt]
      \quad n_6 \frac{dx_6}{dt} = \frac{n_6}{M_6}\bigl[ a_1 x_1 - a_1 x_6 \bigr] + f_6
      \end{multlined}
    \end{aligned}
    \right.
\end{equation}

The key operational distinction lies in liquid level control, which is achieved passively by a fixed-height overflow pipe. When the liquid level in the external liquid region $M_1$ rises above the inlet of this pipe, xenon flows into the dedicated overflow chamber volume $M_6$, from which the liquid return line extracts xenon. Regarding the gas handling, the gaseous xenon volume $M_4$ within the constant-temperature tank is treated as a single unified region due to the absence of any structure isolating the TPC internal gas. Since GI mode is not employed in this run, $M_4$ is replenished solely by evaporation from the liquid xenon surfaces in regions $M_0$ and $M_1$, with the evaporation rate proportional to the respective surface area. The gas return line, equipped with a flow controller that provides the measured flow rate $a_2$, extracts xenon from this unified gas volume and returns it to the circulation loop.

The xenon from both GO and LO is uniformly mixed within the circulation system, then passes through the getter where electronegative impurities are removed. The purified gas re-enters the cycle, repeating the entire process described above. A flow controller on this line indicates the total circulation flow rate $a_{\text{total}}$. Since there is no GI mode in this run, the total flow rate equals the sum of the LI flow rates: $a_{\text{total}} = a_1 + a_2$. Given that the total flow rate and the gas return flow rate $a_2$ are known, the liquid return flow rate is determined by $a_1 = a_{\text{total}} - a_2$.

The most variables and equations are kept consistent with those defined for Run~7 in Eq.~\eqref{eq:impurity_evolution_run7}. However, due to the different regional partitioning in Run~9, both the internal liquid volume $M_0$ and the external liquid reservoir $M_1$ exchange impurities with the unified gaseous volume $M_4$, with the exchange rate proportional to their respective liquid-gas interface areas. While the functional form of the gas-liquid exchange terms remains similar, the expressions have been slightly revised and are defined as follows:

\begin{equation}
\begin{aligned}
T_{0} =  h \cdot A_{\mathrm{TPC}} \cdot (K_H \cdot \frac{\rho_g}{M_{\mathrm{Xe}}} \cdot x_0 - I_{4} )\\[10pt]
T_{1} =  h \cdot A_{\mathrm{Out}} \cdot (K_H \cdot \frac{\rho_g}{M_{\mathrm{Xe}}} \cdot x_1 - I_{4} ).
\end{aligned}
\end{equation}

Furthermore, since only the xenon returning from the gaseous volume $M_4$ and from the overflow chamber $M_6$ is mixed in the circulation loop, the impurity concentration of the gas after purification by the getter is defined as:
\begin{equation}
x_{\mathrm{mix}} = \left( \frac{a_1 \cdot x_6 + a_2 \cdot x_4}{a_1 + a_2} \right) (1 - e).
\end{equation}

The treatment of discontinuities is consistent with that of Run~7, where new $x_i$ parameters are introduced as new starting points.

\subsection{Purification model validation}

We perform a joint fit of the two runs, Run~7 and Run~9. The fit comprises a total of 75 parameters, of which 56 are free parameters related to the initial concentrations, one is a constrained parameter, and the remaining 18 are free parameters associated with the detector. In fitting the model, circulation rates are fixed from flow meter data, while other parameters are either free or constrained by prior measurements. Despite operational interruptions, parameters intrinsic to the detector and purifier are found to remain stable across runs. Several parameters are treated as free parameters to be determined by the fit, with the final results presented in Table~\ref{table:combined_all}:

\begin{itemize}
    \item The initial impurity concentration, $x_{i_0}$, in each volume $i$ at the start of the fitting period. Although these initial concentrations introduce 56 free parameters, they are largely uncorrelated with the intrinsic detector properties and exhibit only weak correlations with other physical parameters. Consequently, their fitted values are not of direct interest in this study.
    
    \item The getter purification efficiency, $e$, is shared and remains consistent between the two runs, as it is the same purifier.
    
    \item The efficiency of the liquid delivery line entering the TPC, $p$, depends on the structural design and is not the same between the two runs.
    
    \item The strength coefficient of the gas-liquid exchange process, $h$, is also shared and remains consistent between the two runs, because the inner and outer vessel designs are consistent, and the cryocooler power and heat leakage show little difference.
    
    \item The outgassing rates for most volumes, namely $f_0$, $f_1$, $f_4$, and $f_5$ for Run~7 or $f_0$, $f_1$, $f_2$, $f_4$, and $f_6$ for Run~9, are related to the detector design and materials. Although these labels represent the same regional definitions, their values are not the same between the two runs. The outgassing rate $f_2$, which corresponds to the cryocooler and related systems, underwent no hardware modifications and therefore remains consistent between the two runs.
    
    \item The effective electron lifetime reduction due to electric field non-uniformity, $\tau_i$, is related to both the electric field strength and the detector design.
\end{itemize}

In contrast, the outgassing rate for the outgassing vessel, $f_3$, is treated as a constrained parameter. This distinction arises because its value is directly informed by independent laboratory measurements conducted at room temperature, which closely matches the operational temperature of this specific vessel. This in situ room-temperature measurement is more reliable because the outgassing material is consistent, it does not rely on extrapolation from cryogenic conditions, and the errors introduced by the measurement method are relatively small. So outgassing in $M_3$ is assigned a Gaussian prior constraint based on these PTFE measurement data.

The parameter estimation in this work is performed using the \texttt{emcee} package~\cite{foreman2013emcee,foreman2019emcee}, an implementation of an affine-invariant ensemble sampler for Markov Chain Monte Carlo (MCMC)~\cite{CHIB20013569}. This method efficiently explores the parameter space by employing an ensemble of walkers that collectively adapt to the geometry of the target posterior distribution. The walkers iteratively sample the parameter space to optimize the parameters, with the goal of converging to the region of maximum likelihood. The total objective function for the fit is defined as the negative log-likelihood, which combines the goodness-of-fit to the data and penalty terms for constrained parameters:

\begin{equation}
\begin{multlined}
\log \mathcal{L}(\theta) =  -\frac{1}{2} \sum_{i=1}^{N} \left[ \frac{(y_i - f(x_i, \theta))^2}{\sigma_i^2} \right] 
\\[10pt] + \sum_{j \in \mathcal{C}} \left[ -\frac{1}{2} \left( \frac{\theta_j - \mu_j}{\sigma_j} \right)^2 \right],
\end{multlined}
\end{equation}
where the first term represents the standard negative half of the $\chi^2$ statistic, which quantifies the goodness of fit between the model predictions, denoted as $f(x_i, \theta)$, and the set of $N$ observed data points, each with its associated uncertainty $y_i \pm \sigma_i$~\cite{mackay2003information}. The second term incorporates Gaussian prior constraints for a designated subset of parameters, such as the outgassing rate $f_3$, where the prior means $\mu_j$ and variances $\sigma_j^2$ are constrained by actual measurements. The MCMC sampler explores the parameter space to maximize the total log-likelihood $\log \mathcal{L}(\theta)$, thereby identifying the parameter set that provides the optimal compromise between fitting the data and adhering to the specified prior constraints.

\begin{table*}[htpb]
\centering
\caption{Best-fit parameters for the Run~7 and Run~9 purification models. 
Parameters $a_1$, $a_2$, and $a_3$ represent the flow rates of liquid circulation, 
gas circulation, and the connected outgassing volume, respectively. 
The definitions of volumes are consistent with those in Section~\ref{sec:model_all}. 
Parameter types and prior constraints are based on independent measurements. 
Best-fit values are reported with their uncertainties. 
The expectation values are derived from room-temperature outgassing rate measurements 
and the calculated low-operating-temperature values, based on material types 
and surface areas of each volume (see Section~\ref{subsec:outgassing}). 
The last two columns present predictions for RELICS-10 and RELICS-50 
using the Run~9 model.}

\label{table:combined_all}
\begin{tabular}{lccccccc}
\toprule
\\[-8pt]
\textbf{Parameter} & \textbf{Type} & \multicolumn{2}{c}{\textbf{Run 7}} & \multicolumn{2}{c}{\textbf{Run 9}} & \multicolumn{2}{c}{\textbf{Prediction}} \\[2pt]

& & \textbf{Best-Fit} & \textbf{Expectation} & \textbf{Best-Fit} & \textbf{Expectation} & \textbf{\quad RELICS-10} & \textbf{\quad RELICS-50} \\[2pt]
\toprule
\\[-8pt]
Circulation Mode & -- & \multicolumn{2}{c}{MI-MO}& \multicolumn{2}{c}{LI-MO or LI-GO} & \multicolumn{2}{c}{LI-MO}\\[2pt]
Main Material & -- & \multicolumn{2}{c}{PTFE}& \multicolumn{2}{c}{PEEK} & \multicolumn{2}{c}{PTFE}\\[2pt]

$a_1$ [SL/min] & Fixed & \multicolumn{2}{c}{\multirow{3}{*}{Fig.~\ref{fig:operation_run7}}} & \multicolumn{2}{c}{\multirow{2}{*}{Fig.~\ref{fig:operation_run9}}} & 15 & 40 \\[2pt]
$a_2$ [SL/min] & Fixed & & & & & 5 & 10 \\[2pt]
$a_3$ [SL/min] & Fixed & & &-- & -- & -- & -- \\[2pt]

$M_0$ [kg] & Fixed & 1 & -- & 1 & -- & $10 \pm 10\%$ & $50 \pm 10\%$ \\[2pt]
$M_1$ [kg] & Fixed & 8 & -- & 8 & -- & $20 \pm 10\%$ & $40 \pm 10\%$ \\[2pt]
$M_2$ [g] & Fixed & 300 & -- & 300 & -- & $300 \pm 10\%$ & $800 \pm 10\%$ \\[2pt]
$M_3$ [g] & Fixed & 80 & -- & -- & -- & -- & --\\[2pt]
$M_4$ [g] & Fixed & 45 & -- & 150 & -- & $200 \pm 10\%$ & $600 \pm 10\%$ \\[2pt]
$M_5$ [g] & Fixed & 75 & -- & -- & -- & -- & -- \\[2pt]
$M_6$ [kg] & Fixed & -- & -- & 2 & -- & $3 \pm 10\%$ & $5 \pm 10\%$ \\[2pt]
$n_i$ [mol] & Fixed & -- & -- & -- & -- & -- & -- \\[2pt]

\multirow{2}{*}{$k_{E}$ [L/(mol·s)]} & \multirow{2}{*}{Fixed} 
& $1.4 \times 10^{11}$ (E = 166V) & -- & $1.3 \times 10^{11}$ (E = 200V) & -- 
& \multicolumn{2}{c}{\multirow{2}{*}{$1.3 \times 10^{11}$ (E = 200V)}}\\[2pt]
&  & $1.1 \times 10^{11}$ (E = 333V) & --  & $1.0 \times 10^{11}$ (E = 500V) & -- \\[2pt]

$e$ & Free & $0.98 \pm 0.02$ & -- & $0.98 \pm 0.02$ & -- & \multicolumn{2}{c}{$0.95\pm 0.05$} \\[2pt]
$p$ & Free & $0.07 \pm 0.01$ & -- & $0.67 \pm 0.05$ & -- & \multicolumn{2}{c}{$0.67\pm 0.1$} \\[2pt]
$h $ [m/s] & Free & $2.0^{+14.2}_{-1.3} \times 10^{-6}$  & -- & $2.0^{+14.2}_{-1.3} \times 10^{-6}$  & -- & \multicolumn{2}{c}{$2.0^{+14.2}_{-1.3} \times 10^{-6}$ } \\[2pt]

$f_0$ [ppb·mol/h] & Free & $1.9 \pm 0.2$ & $5.6$ & $3.2 \pm 0.9$ & $4.6$ & $4.49\pm 1.1$ & $16.0\pm 3.6$ \\[2pt]
$f_1$ [ppb·mol/h] & Free & $11.4 \pm 1.7$ & $34.3$ & $3.4 \pm 1.4$ & $9.1$ & $ 5.7\pm 1.2$ & $22.2\pm 4.5$ \\[2pt]
$f_2$ [ppb·mol/h] & Free & $0.02 \pm 0.17$ & $\sim 0$ & $0.02 \pm 0.17$ & $\sim 0$ & \multicolumn{2}{c}{$\sim 0$} \\[2pt]
$f_3$ [ppb·mol/h] & Constrained & $163 \pm 17$ & $167 \pm 11$ &-- & --& --& --\\[2pt]
$f_4$ [ppb·mol/h] & Free & $7.0 \pm 7.5$ & $3.4$ & $8.9 \pm 1.5$ & $34$ & $34.8\pm 13.4$ & $123.1\pm 43.9$ \\[2pt]
$f_5$ [ppb·mol/h] & Free & $18.4 \pm 12.4$ & $25.6$ &-- &-- &-- & --\\[2pt]
$f_6$ [ppb·mol/h] & Free &-- &-- & $0.8 \pm 0.5$ & $\sim 0$ & \multicolumn{2}{c}{$\sim 0$} \\[2pt]
\multirow{2}{*}{$\tau_{E}$ [$\mu$s]} & \multirow{2}{*}{Free} 
& $235 \pm 469$ (E = 166V) & -- & $54 \pm 3$ (E = 200V) & -- 
& \multicolumn{2}{c}{\multirow{2}{*}{$\infty$ }}\\[2pt]
&  & $395 \pm 252$ (E = 333V) & --  & $122 \pm 12$ (E = 500V) & -- \\[2pt]

best elife [$\mu$s] & -- & $\sim 150$ & -- & $\sim 120$ & $\sim 700$ & $\sim 2000$ & $\sim 1400$ \\[2pt]

\bottomrule
\end{tabular}
\end{table*}

\begin{figure*}[htbp]
    \centering
    \includegraphics[width=1\linewidth]{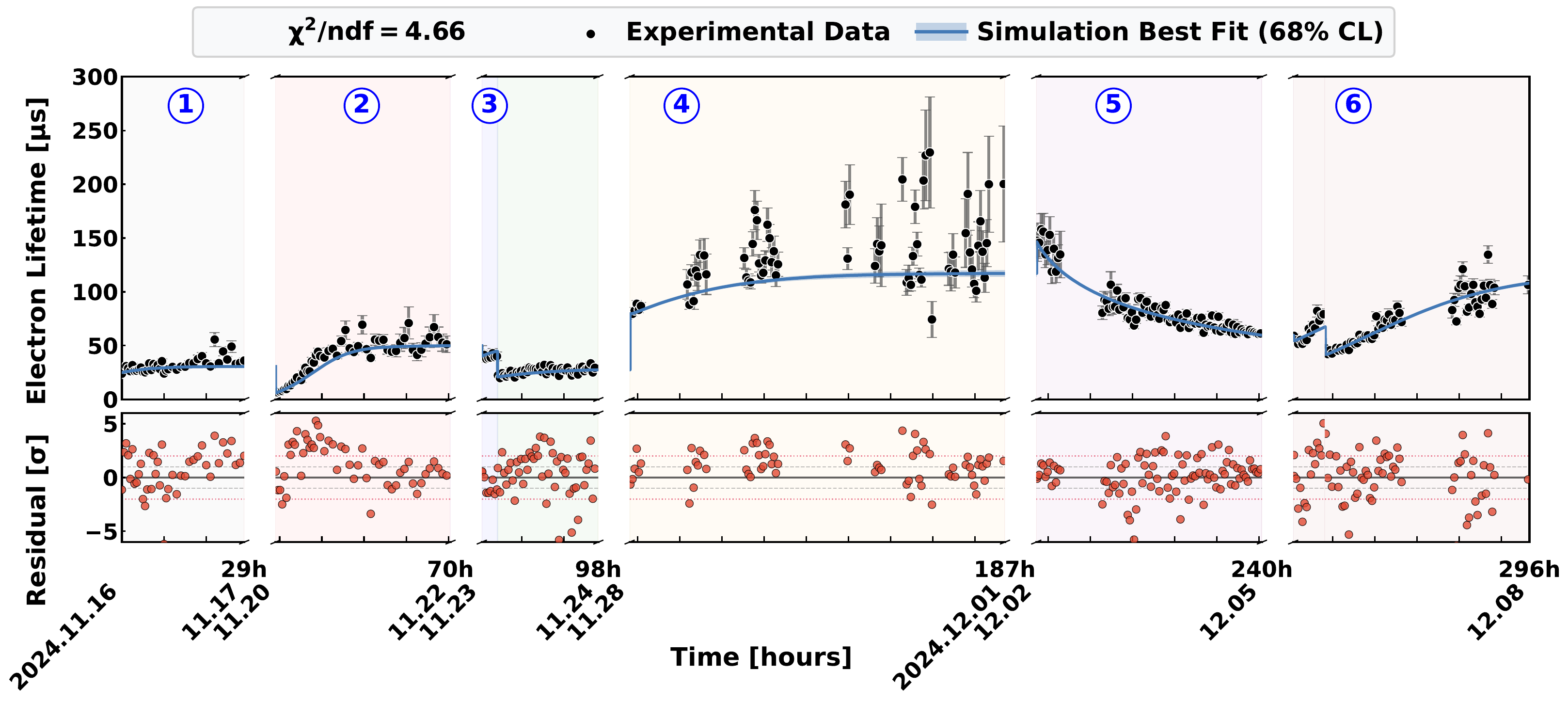}
    \caption{The measured electron lifetime data and the fitted purification model for the Run-7 prototype are shown in the figure. The black dots represent the measured values, while the blue lines indicate the predictions from different walkers in the MCMC fitting. The blue band in the figure denotes the 68\% confidence interval. The left vertical axis shows the electron lifetime. The subfigure below shows the residual distribution of each data point, calculated as \(\frac{\text{Data} - \text{Model}}{\sigma}\). The colored shaded regions represent different time windows corresponding to different operations during the Run~7.}
    \label{fig:run_7_elife}
    
    \vspace{\floatsep}
    
    \includegraphics[width=1\linewidth]{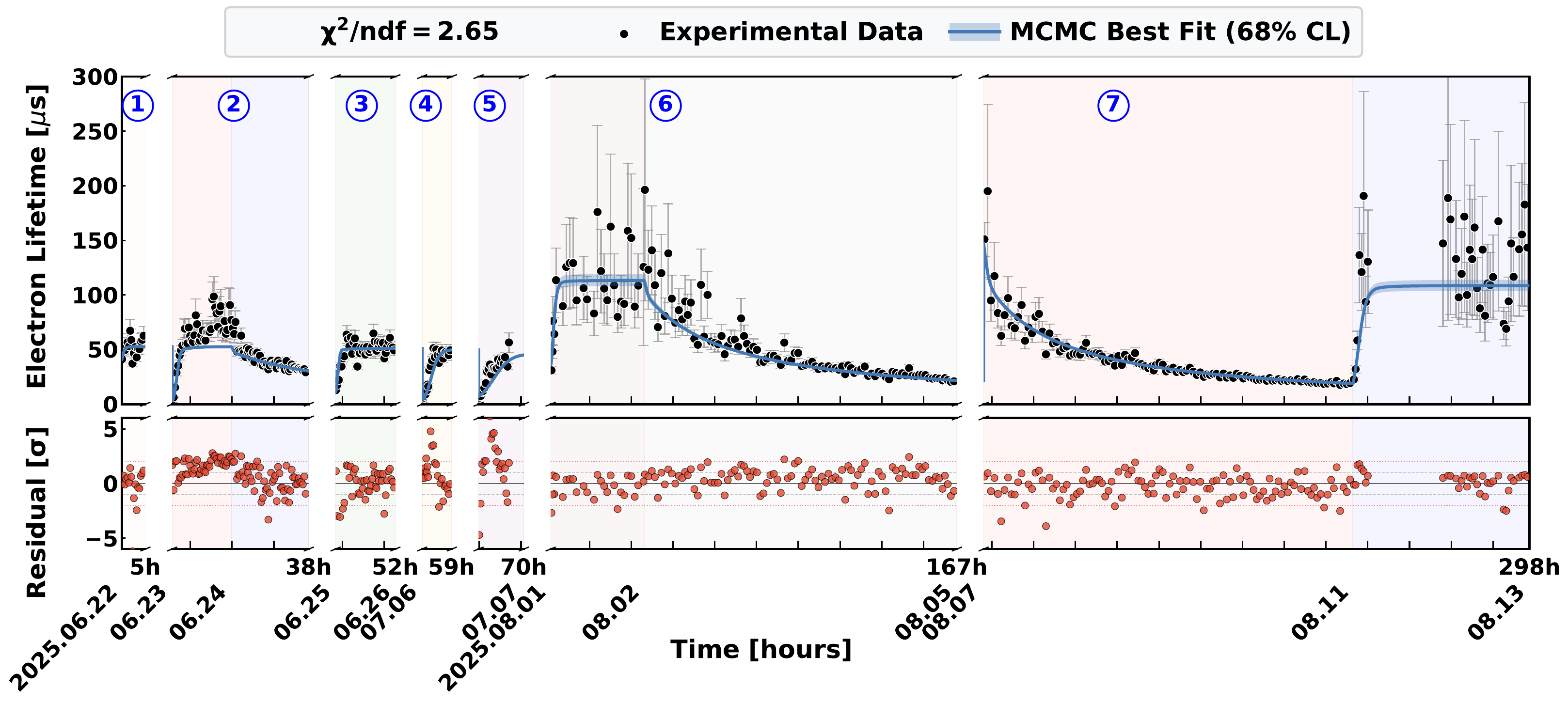}
    \caption{The measured electron lifetime data and the fitted purification model for the Run-9 prototype are shown in the figure. The black dots represent the measured values, while the blue lines indicate the predictions from different walkers in the MCMC fitting. The blue band in the figure denotes the 68\% confidence interval. The left vertical axis shows the electron lifetime. The subfigure below shows the residual distribution of each data point, calculated as \(\frac{\text{Data} - \text{Model}}{\sigma}\). The colored shaded regions represent different time windows corresponding to different operations during the Run~9.}
    \label{fig:run_9_elife}
\end{figure*}

For each run, we employ 1000 walkers, discarding the first 100 steps as burn-in. The chain length is extended until convergence is achieved. The convergence is checked every 100 consecutive steps by examining the change in the chi-square value. If the best-fit chi-square value shows a change of less than 1 compared to that of the previous 100 steps, i.e., $\Delta \chi^2 < 1$, the best-fit values are considered to be reached. Further increasing the number of steps does not significantly change the parameters, nor does it substantially improve the model fit to the data.

The purification models developed for the Run~7 and Run~9 configurations, as detailed in Sections~\ref{sec:run7_model_quaction} and~\ref{sec:run9_model_equation}, are applied to fit the experimentally measured electron lifetime evolution data from Section~\ref{sec:prototype_campaign}. To achieve better parameter constraints, these two models are jointly fitted, maintaining consistency for parameters that are shared or similar between the two runs. Using the MCMC framework described in the previous section, we simultaneously constrain all parameters for both runs. All data points at every time stamp from both campaigns are included in the fit with equal weight; no time periods are down-weighted or excluded. 

The fitting process yields the optimal evolution curves for the electron lifetime, providing a direct comparison between the model predictions and the experimental data, as illustrated in Figs.~\ref{fig:run_7_elife} and~\ref{fig:run_9_elife}. The \(\chi^2/\text{ndf}\) values of the fitting results are provided, where ndf (Number of Degrees of Freedom) is obtained by subtracting the number of free parameters from the number of data points. In addition, the residual distributions across all data points are presented, calculated as:
\begin{equation}
\text{residual} = \frac{\text{Data} - \text{Model}}{\sigma}.
\end{equation}

From the residual distributions of the two fits, most data points fall within the $3\sigma$ band of the model, indicating that the purification model captures the characteristics of the data variations. Some deviations are observed, such as the rising phase of Phase~2 shown in Fig.~\ref{fig:run_9_elife}. Although no changes were made to the hardware equipment or the detector operating voltage, it is suspected that this deviation is due to some variation in the electric field distortion parameters. This will be discussed in detail later.

This joint fitting exercise serves two purposes: it validates the model's capability to describe the impurity dynamics under two different detector designs, namely the diving bell design for Run~7 and the overflow chamber design for Run~9, and it extracts the key physical parameters governing the purification process for both configurations. The corresponding best-fit values and their uncertainties for all model parameters from both runs are quantitatively determined and summarized in Table~\ref{table:combined_all}. In addition, the table also provides information on the assumed masses of each region in the model, the circulation modes, and the parameter types, among other details.

The masses and circulation flow rates, which are determined by the detector design and flow controllers respectively, are input into the purification model as known quantities. The Expectation values listed in the table correspond to the independently measured outgassing rates of the materials in each volume, obtained from room-temperature measurements and subsequently extrapolated to cryogenic conditions using temperature scaling. The Best-Fit values, on the other hand, are the outgassing rates derived from the fitting procedure. The discrepancies between these two sets of values are discussed in detail in Section~\ref{subsec:outgassing}. Furthermore, several parameters closely related to the operation are also discussed in detail in this section, including the getter purification efficiency, which remains at a high level, the improvement in the liquid delivery line efficiency, the influence of the electric field on purity measurements, and the handling of special cases during operation.

\subsubsection{High getter efficiency}

First, the purifier used in our system possesses a high impurity removal capacity. To verify this, in the Run~7 design, a room-temperature chamber containing Teflon was placed upstream of the purifier to test the purification efficiency of the getter. A direct comparison of the purity plateaus under different configurations demonstrates that both reach an identical final purity level. 

\begin{figure}[htbp]
    \centering
    \includegraphics[width=0.99\linewidth]{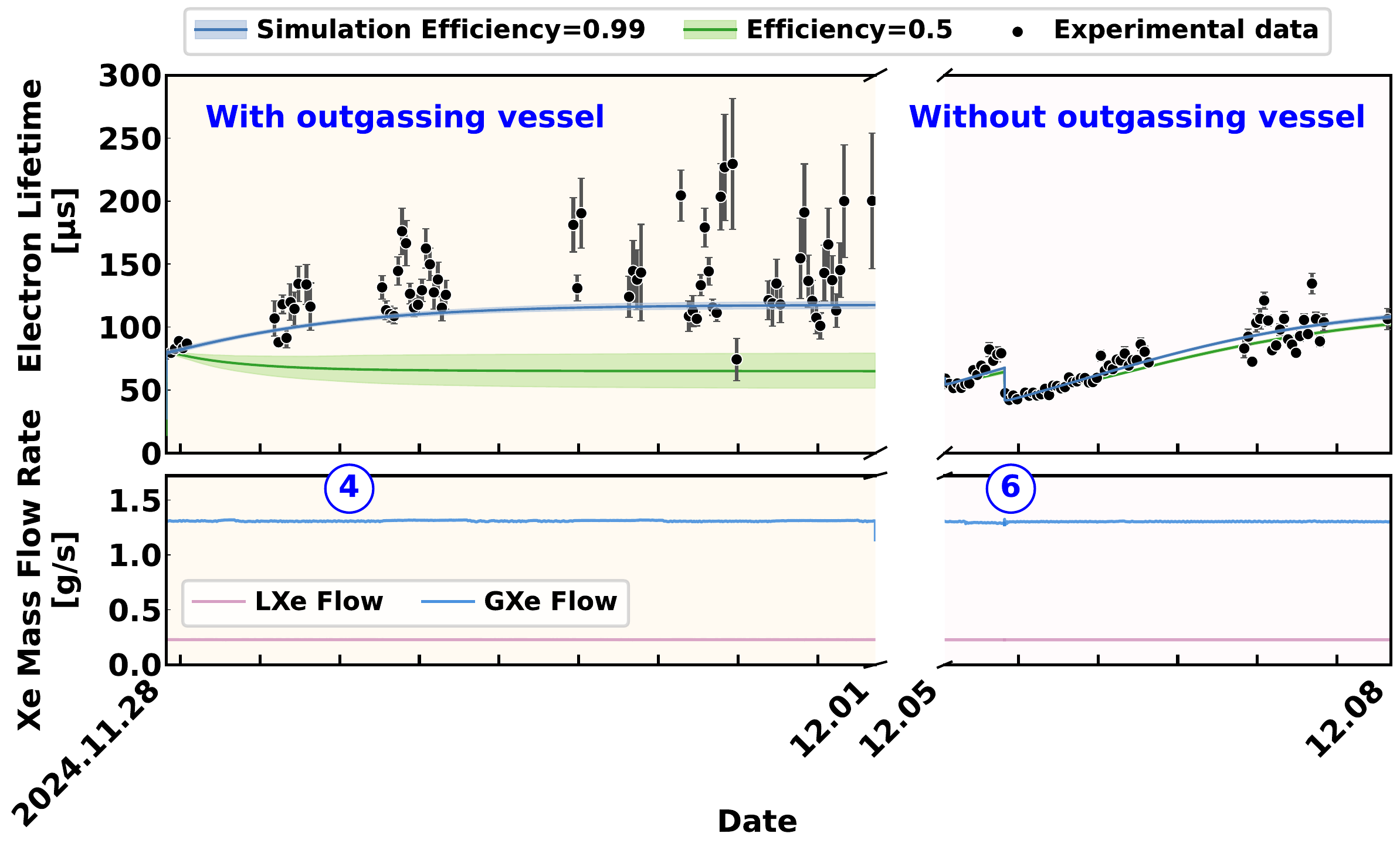}
    \caption{Selected data points (Phase~4, Phase~6) compare system performance with and without the outgassing vessel to determine the getter efficiency. The upper panel shows data and model fits; the lower panel shows circulation flow rates. The blue curve is the best fit, and the green curve shows the prediction for 50\% efficiency. The black points represent the experimental data. The colored shaded regions correspond to different operation time windows in Run~7.}
    \label{fig:getter}
\end{figure}

For example, one configuration connects the outgassing vessel during Phase~2, while another has it disconnected during the period of Phase~1. A similar comparison can be made between later periods, specifically Phase~4 and Phase~6. This observation can, in principle, be explained by two distinct mechanisms. The first mechanism is that the getter possesses an exceptionally high purification efficiency, enabling it to remove the vast majority of introduced impurities regardless of the vessel's connection status. Alternatively, the second possible mechanism is that the outgassing vessel itself contributes a negligible amount of impurities because its intrinsic outgassing rate is extremely low. Crucially, independently measured data show a high outgassing rate for PTFE at room temperature, which directly contradicts the assumption of a low outgassing rate from the vessel. Therefore, the high purification efficiency of the getter remains the only consistent explanation. As shown in Fig.~\ref{fig:getter}, data points Phase~4 and Phase~6 are analyzed under constant circulation flow rate and electric field, with the only operational difference being whether the vessel is connected or disconnected. That both conditions yield similar high purity levels indicates that the getter maintains high purification efficiency. This interpretation is further reinforced by a counterfactual scenario: if the purification efficiency were lower, as modeled by the green curve, the purity for data point Phase~4 would decrease significantly, which is not observed. This sufficiently demonstrates that the getter has an extremely high purification efficiency.

\subsubsection{Improvement of liquid delivery efficiency}

\begin{figure}[htpb]
    \centering
    \includegraphics[width=1\linewidth]{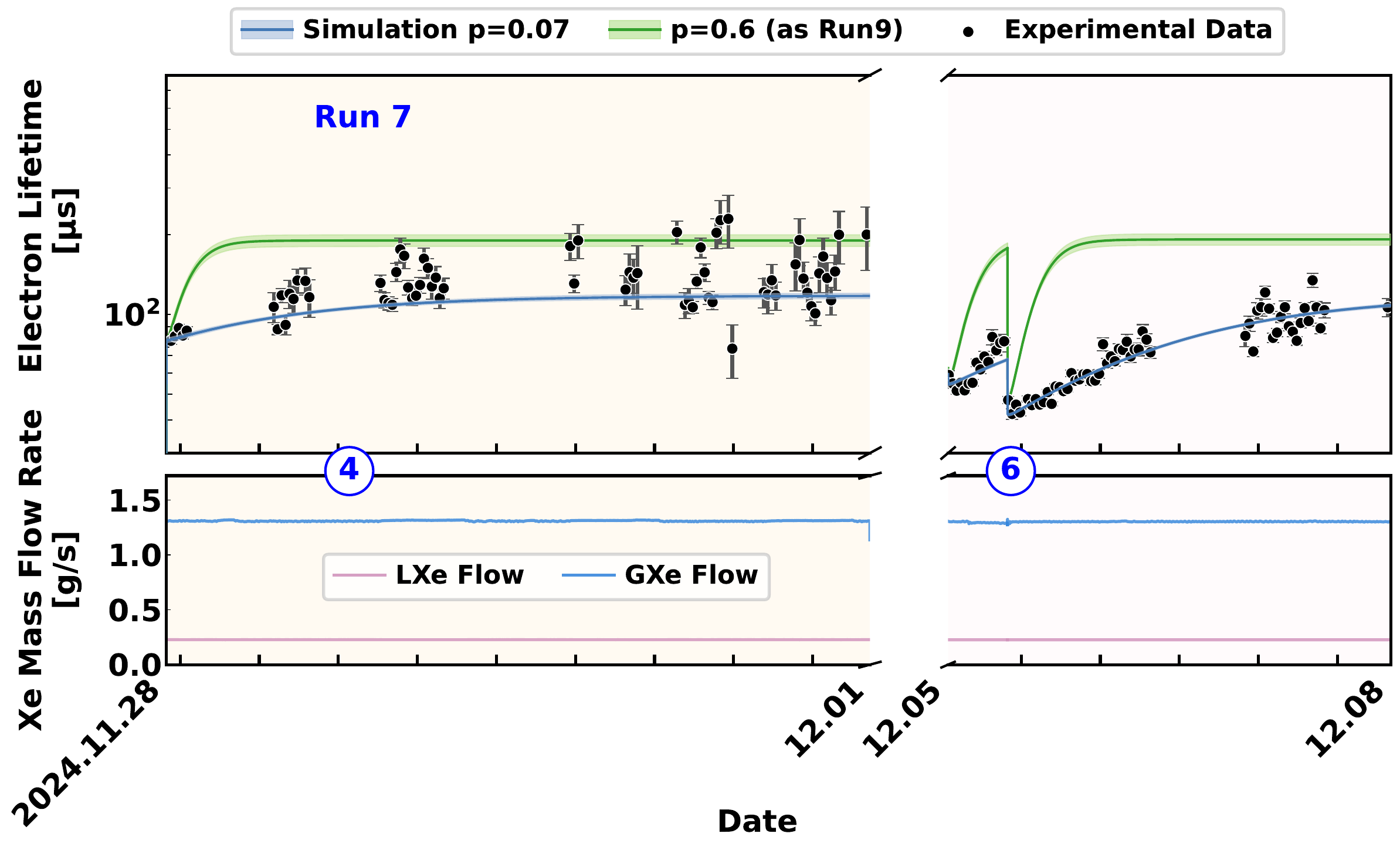}
   \caption{Selected data points Phase~4 and Phase~6 illustrate the effect of imperfect sealing in the delivery line. The upper panel shows data and model fits; the lower panel shows circulation flow rates. The blue curve is the best fit, and the green curve shows the prediction if delivery efficiency were 0.6. The black points represent the experimental data. The colored shaded regions correspond to different operation time windows in Run~7.}
    \label{fig:pvalue}
\end{figure}

Even with this high purification efficiency of the getter, the data further show that the purity ascent phases require approximately 40 hours to complete, as observed during the periods of data points such as Phase~2 and Phase~6. In contrast, the purity decrease driven solely by detector material outgassing proceeds very slowly, as seen during the period of data point Phase~4. This contrast indicates an intrinsically low outgassing rate from the main detector materials at the operational cryogenic temperature. Therefore, the protracted ascent time cannot be attributed to a high intrinsic outgassing load from the detector materials. Instead, one plausible explanation attributes the protracted ascent time to a specific design factor: a significant fraction of the purified LXe bypasses the internal target volume $M_0$ due to sealing inefficiencies in the flexible hose of the delivery line and enters the external reservoir $M_1$. Quantitatively, only about 7\% of the purified liquid is delivered into $M_0$. This substantial bypass flow markedly reduces the effective purification rate within the core detection volume, explaining the slow ascent. As shown in Fig.~\ref{fig:pvalue}, the ascent phases from data points Phase~4 and Phase~5 are selected to illustrate this point. The key parameter $p$ corresponds to the delivery efficiency of the purified liquid into $M_0$. The analysis suggests that if the sealing integrity of the delivery line were higher, meaning a higher value of the efficiency parameter $p$, the ascent rate would be faster and a higher purity plateau should be attainable, as indicated by the green curve in the figure. However, the measured data do not exhibit this faster ascent, indicating that the current configuration does not achieve such a high delivery efficiency. This observed limitation implies that the effective gas–liquid exchange term in the current system is also relatively small. Consequently, aside from the circulation flow itself, there is no strong additional purification effect significantly enhancing the ascent rate, which is consistent with the observed slow purification dynamics.

For comparison, analysis of the Run~9 data confirms significant operational improvements. The adoption of a VCR-reinforced connection for the liquid delivery line enhanced its sealing integrity. This modification is directly evidenced by the purification ascent phases observed during the periods from data point Phase~1 to Phase~6. The time required to reach a high purity plateau was reduced from nearly two days to approximately four hours. These results demonstrate a substantial improvement in the liquid delivery efficiency, with the estimated parameter \(p\) increasing from the previous level of about 7\% to approximately 65\%. This conclusion can be drawn from the comparison of the ascent phases in Fig.~\ref{fig:run_7_elife} and Fig.~\ref{fig:run_9_elife}, while Fig.~\ref{fig:pvalue} substitutes the high liquid delivery efficiency of Run~9 into the simulation data, providing a direct visual comparison of how the detector design upgrade effectively improves the \(p\) value between the two runs.

\subsubsection{Speculation on cold-head ice formation.}

In the operation of Run~7, the early period (Phase~1--Phase~3) and the later period (Phase~4 -Phase~6) exhibit different purity behaviors, with lower purity levels in the early period and higher levels later. This change is difficult to explain solely by variations in the circulation flow rate, since the total circulation flow rate did not increase, and the liquid circulation flow rate was continuously decreasing. To maintain consistency throughout the operation, the existing parameters are insufficient to account for this observation, and the introduction of a new degree of freedom is required.

In particular, monitoring of the cryocooler cold-head temperature combined with its operational status suggests the possible formation of impurities may have frozen on the cold head during a period encompassing data points Phase~1, Phase~2 and Phase~3. A series of diagnostic adjustments was subsequently implemented. These adjustments appear to have resolved the issue, leading to the sublimation of the ice and recovery of normal purity levels, as observed in the later data points Phase~4, Phase~5 and Phase~6. One interpretation is that this ice layer introduced a significant outgassing source, which can be modeled as an additional outgassing rate variable $f_{\text{coldhead}}$ in the $M_2$ region of the Run~7 model. The fitted value of this parameter in the purification model is approximately 129 ppb mol/h. The presence of such a source could explain the suppressed purity observed during high-flow-rate operations in the earlier period. 

\begin{figure}[htbp]
    \centering
    \includegraphics[width=0.99\linewidth]{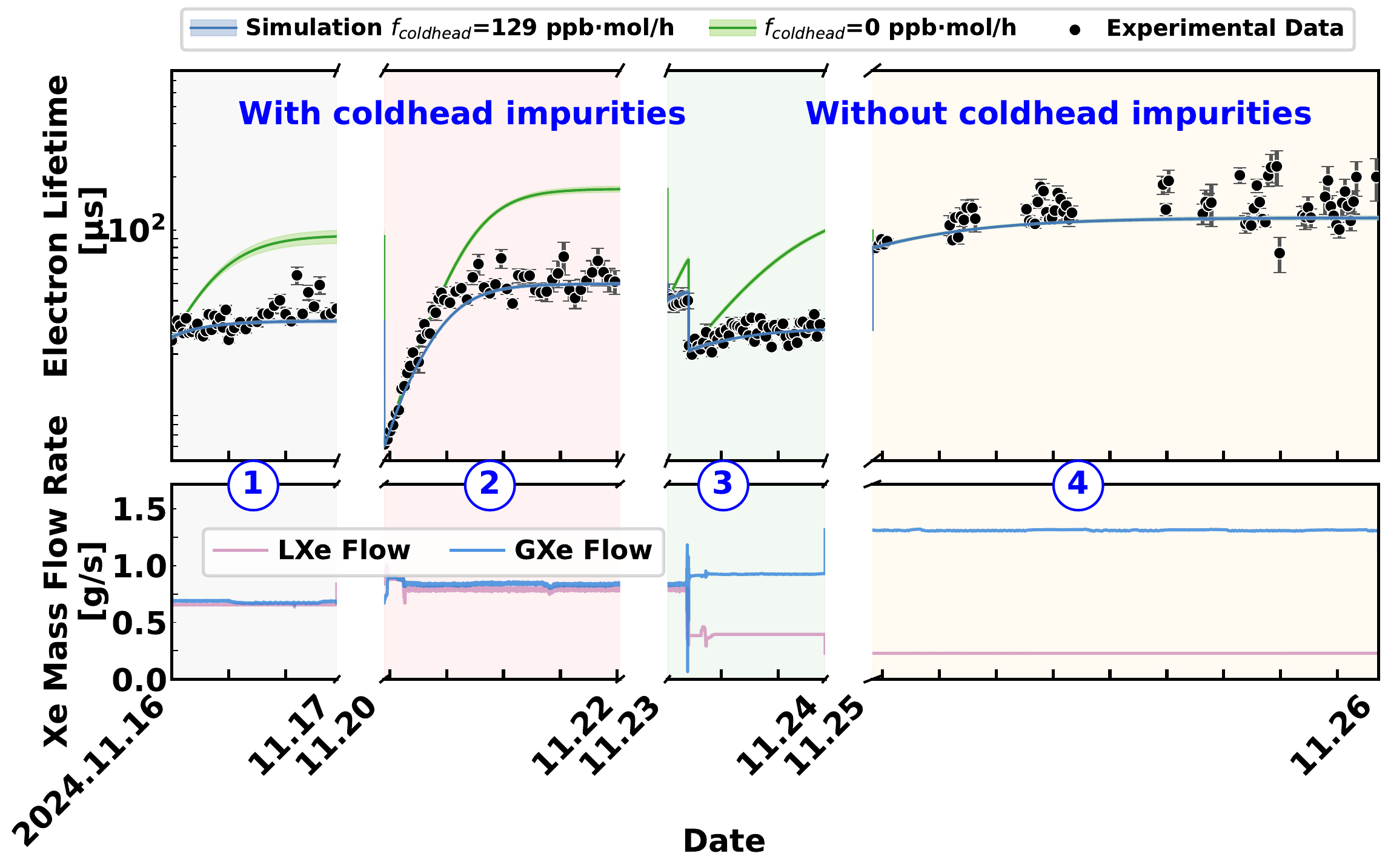}
    \caption{Selected data points (Phase~1--Phase~4) illustrating the additional outgassing source from the cold head. The upper panel shows the experimental data and model fits, and the lower panel shows the circulation flow rates. The blue curve represents the best fit including the additional outgassing source, while the green curve assumes its absence. The black points represent the experimental data. The colored shaded regions correspond to different operation time windows in Run~7.}
    \label{fig:coldhead}
\end{figure}

To assess this interpretation, a counterfactual scenario is considered. If this postulated factor were absent, the system might have exhibited a tendency toward higher purity at high flow rates. This expected behavior is illustrated by the green curve in Fig.~\ref{fig:coldhead}, particularly at data points such as Phase~1, Phase~2 and Phase~3. This expectation gains support from a direct comparison of the data: the circulation flow rates for these earlier points are all higher than those for Phase~4, Phase~5 and Phase~6, yet the corresponding electron lifetimes were lower. This observed pattern is consistent with the hypothesized presence of an outgassing source that existed earlier but later disappeared.

\subsubsection{field-distortion effect}

As discussed above, in Run~9, the connection between the detector and the liquid delivery line was improved. Compared with Run~7, the effectiveness of this improvement is enhanced by approximately one order of magnitude, and the duration of the purity rising phase is significantly shortened. Moreover, while the material outgassing rates show no significant difference between the two runs, the purity plateau in Run~9 remains at a relatively lower level. We attribute this observation to the non-uniformity of the electric field.

\begin{figure}[htbp]
    \centering
    \includegraphics[width=0.99\linewidth]{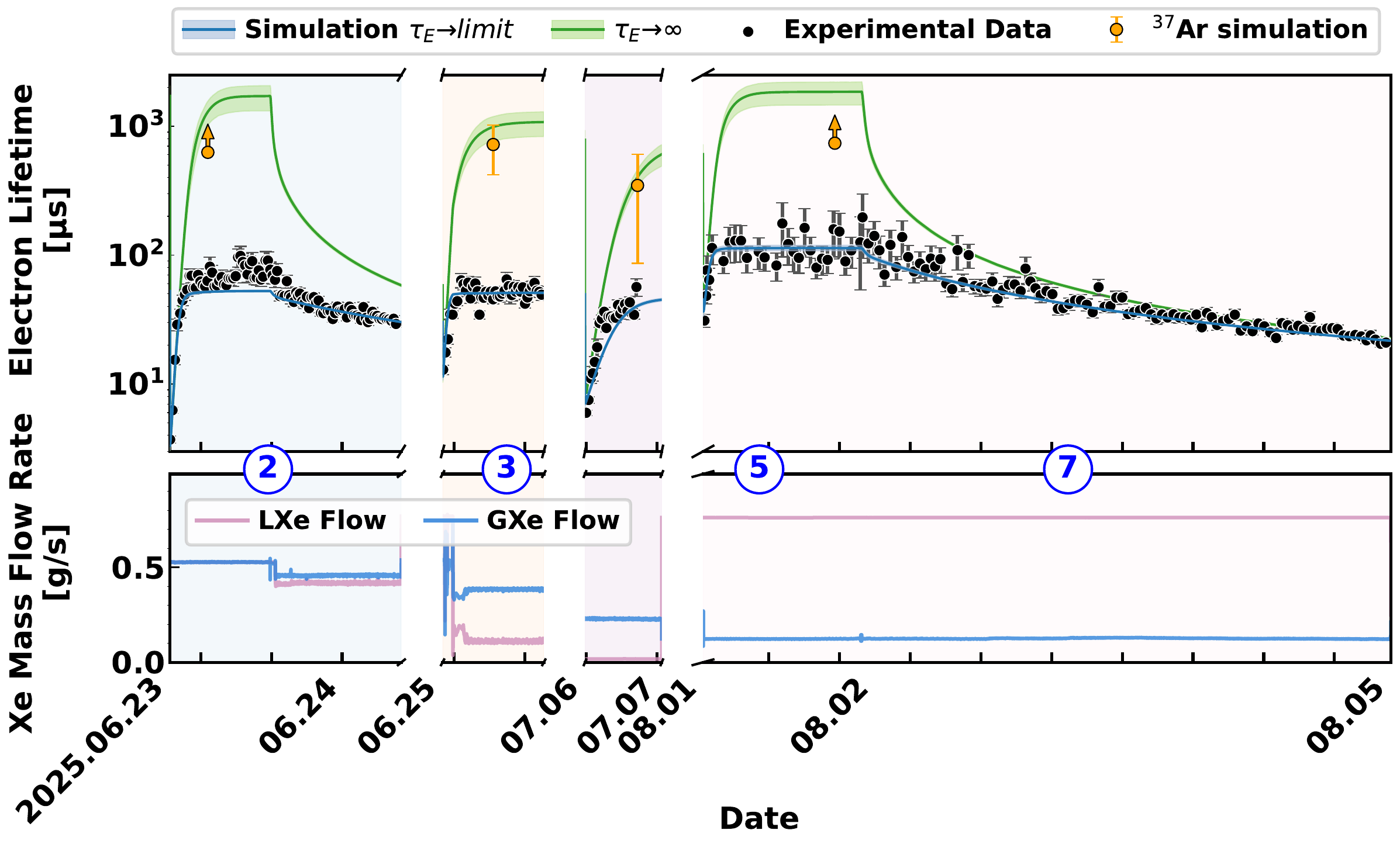}
    \caption{Data points Phase~2, Phase~3, Phase~5 and Phase~7 illustrate the electron lifetime limit imposed by electric field inhomogeneity. The upper panel shows data and model fits; the lower panel shows circulation flow rates. The blue curve represents the measured data, and the green curve shows the predicted electron lifetime if determined solely by impurity concentration without field inhomogeneity. The colored shaded regions correspond to different operation time windows in Run~9.}
    \label{fig:efield}
\end{figure}

Under conditions characterized by high purification efficiency of the getter and the slow purity decrease rates observed during the periods corresponding to data points Phase~2, Phase~6 and Phase~7 directly point to a low intrinsic outgassing rate from the detector materials at cryogenic temperatures. As shown by the green curve in Fig.~\ref{fig:efield}, the model predicts a corresponding electron lifetime plateau around 1 ms under these conditions and the predicted ascent trend during purification phases is consistent with the data. However, the measured electron lifetime plateau stabilizes at a much lower level than expected, and this level remains unchanged even when circulation flow rates are varied. Given that the Run7 configuration does not exhibit a strong limitation, a comparison of the detector structures is instructive. In the Run9 design, the aluminum and PEEK field-shaping rings are in direct contact with the LXe volume, while in the Run7 configuration, a PTFE insulator layer separates the copper field rings from the LXe. Both designs maintain consistent spacing and width of the field-shaping rings, and electric field simulations show similar behavior, with field distortions at the edges and slight improvements in the central region. However, based on data comparison, the drift time in Run7 varies almost uniformly with radius, indicating better internal field performance. In Run9, the drift times at the edges are not uniform, and distortions also exist in the central region. A possible explanation is that the monolithic PTFE structure in Run7 may facilitate charge adsorption, allowing positive charge to accumulate at the edges of the PTFE. This phenomenon results in a more uniform radial electric field. In Run9, the alternating Al and PEEK structure prevents charge accumulation, so the distortions are not effectively mitigated. Consequently, the electric field effects are more severe in Run9 than in Run7, manifesting as a flow-rate-independent upper limit on the measurable electron lifetime $\tau_{e}$ within the TPC's active region. This effect cannot be quantitatively simulated and is therefore treated as a free parameter in the fitting process.

To test this hypothesis and mitigate the measurement limitation, a higher drift electric field was applied starting after the period of data point Phase~6. This intervention raised the observed electron lifetime plateau from approximately 50 $\mu$s to about 120 $\mu$s, confirming that field uniformity was a limiting factor. Crucially, data taken subsequently on both data point Phase~6 and Phase~7 under different flow rates consistently reached this new, identical plateau level. This provides definitive evidence that electric field distortion, not impurity dynamics, constitutes a primary factor limiting the apparent electron lifetime in the Run~9 configuration. 

To further investigate the impact of electric field distortion, the $^{37}$Ar signal was sought in the detector~\cite{aprile2023low,be2013table}. Its energy region is significantly lower than that of $^{83\text{m}}$Kr. According to the electron yield versus electric field curves in NEST~\cite{szydagis2025review,akimov2014experimental}, the variation in electron yield induced by electric field changes is smaller in the low-energy region. Therefore, in the $^{37}$Ar energy region, we consider the upper limit on the electron lifetime due to electric field non-uniformity to be higher than that for $^{83\text{m}}$Kr. Thus, the two monoenergetic signals at different energy regions in the detector can be used for cross-checking. During the purity plateau period, $^{37}$Ar data are accumulated over an extended period. The electron lifetime obtained from these data is then used to infer the electron lifetime in the absence of electric field distortion, by applying the slope derived from NEST simulations. The result, shown as the yellow data point in Fig.~\ref{fig:efield}, is consistent with the model prediction, both indicating that the electron lifetime should exceed 700~$\mu$s, or even higher.

\subsubsection{parameter correlation}

\begin{figure}[htpb]
    \centering
    \includegraphics[width=1\linewidth]{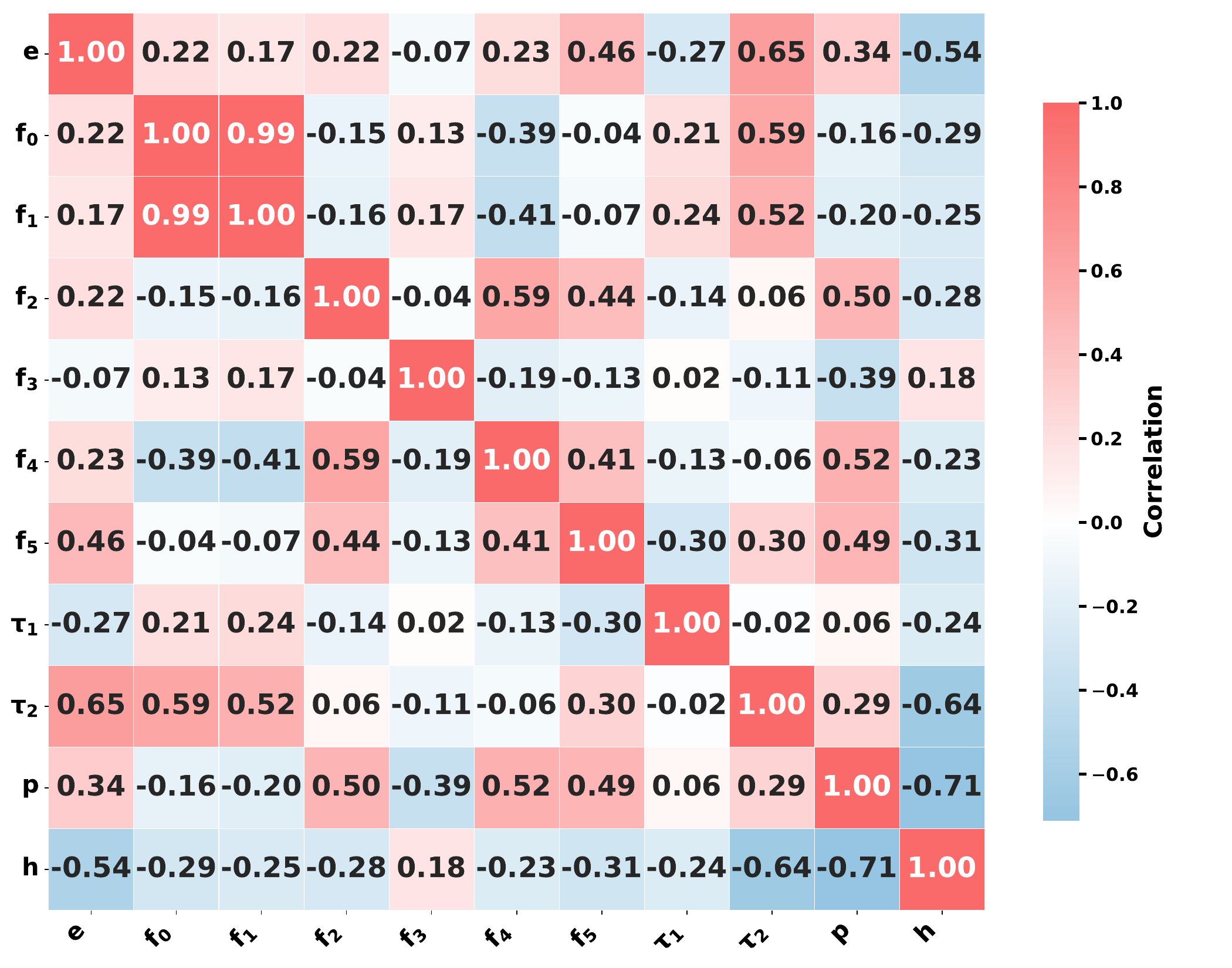}
   \caption{ Correlations among detector-related parameters in the Run~7 fitting results. Red indicates a positive correlation, and blue indicates a negative correlation.}
    \label{fig:run7_correlation}
\end{figure}

\begin{figure}[htpb]
    \centering
    \includegraphics[width=1\linewidth]{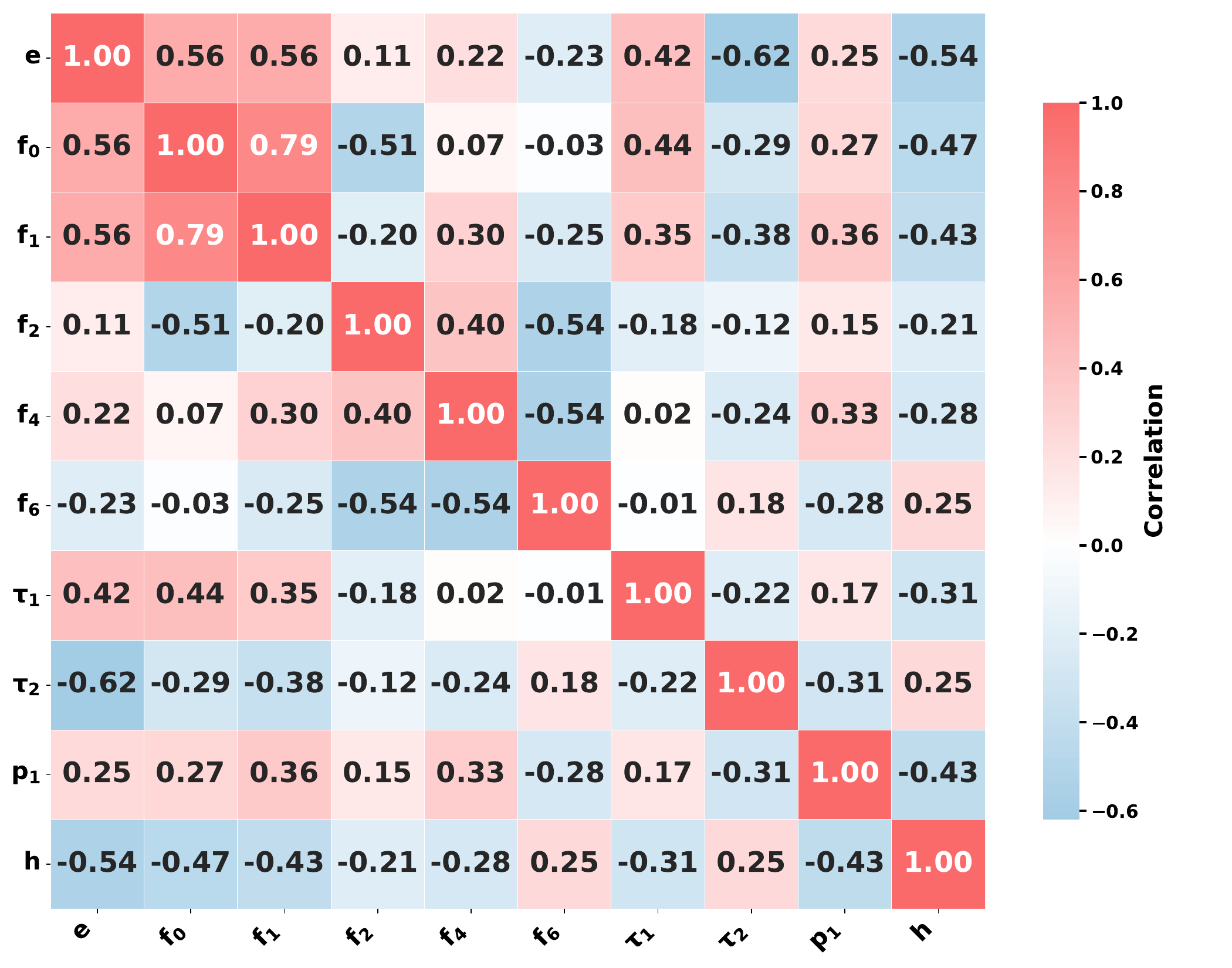}
   \caption{ Correlations among detector-related parameters in the Run~9 fitting results. Red indicates a positive correlation, and blue indicates a negative correlation.}
    \label{fig:run9_correlation}
\end{figure}

In the joint fitting of the two runs, the parameters $e$, $h$, and $f_2$ are kept consistent, while the remaining parameters are treated as uncorrelated free parameters specific to each run. The parameter correlations for Run~7 and Run~9 are shown in Fig.~\ref{fig:run7_correlation} and Fig.~\ref{fig:run9_correlation}, respectively. The 56 initial concentration parameters $x_i$, which are unrelated to detector performance and are not of primary interest, are omitted from the figures.

The parameter correlations in the Run~7 fitting results are shown in Fig.~\ref{fig:run7_correlation}. The parameters $f_0$ and $f_1$ represent the outgassing rates of regions dominated by PTFE materials at liquid xenon temperature, while $f_1$ also includes contributions from cables and other materials. The surface area ratio of these two regions can be calculated from the detector design drawings and is therefore taken into account in the model fitting. Since the outgassing rate associated with the PTFE surface area is overwhelmingly dominant, the contributions from other materials in $M_1$ are relatively small, leading to a strong correlation between $f_0$ and $f_1$. Due to the exchange of xenon and impurities between different regions, certain correlations also exist among the outgassing rates of regions connected by such exchange. Moreover, the stronger the exchange between regions, the stronger the correlation between their outgassing rates. Similar behavior is also observed in the Run~9 parameter correlations, as shown in Fig.~\ref{fig:run9_correlation}.

Meanwhile, based on the definitions within the purification model, the getter purification efficiency $e$ and the liquid delivery efficiency $p$ of the TPC, the product of which represents the purification efficiency of the circulation process, together with the gas-liquid exchange term $h$, are the key parameters that determine the rate of the purity rising phase. Consequently, they exhibit an overall negative correlation. The final purity plateau, on the other hand, is determined by the outgassing rate and the purification effectiveness. For a given purity level, these two factors exhibit a positive correlation. Assuming a single independent volume, the electron lifetime can be expressed as:
\begin{equation}
\begin{split}
\tau_e &= \frac{1}{k_E \cdot I_{[O_2]}} \\
&= \frac{1}{k_E \left( \left( I - \frac{M}{p \cdot a \cdot e} \cdot \frac{f}{V} \right) e^{-\frac{p \cdot a \cdot e}{M} \cdot t} + \frac{M}{p \cdot a \cdot e} \cdot \frac{f}{V} - T \right)},
\end{split}
\end{equation}
where the variables are defined consistently with the model assumptions. The observed correlations among the parameters are consistent with the processes described in the purification model.

\subsection{Material Outgassing Characterization}
\label{subsec:outgassing}

Establishing a robust purification model for the RELICS program requires a precise quantification of the impurity source terms inherent to the detector geometry. Consequently, we characterized the specific outgassing rates of the primary construction materials used in the TPC and cryostat, namely stainless steel, PTFE, PEEK, polyimide (Kapton), and PMT bases. These measurements provide the fundamental input parameters necessary to predict the impurity evolution in both the prototype and the RELICS detectors.

The outgassing rates were determined at room temperature using the Rate-of-Rise (ROR) method, in which the specific gas load is inferred from the pressure increase in an isolated stainless steel vessel as a function of time~\cite{OHanlon2003,Fedchak2021}. 
Simultaneously, a residual gas analyzer (RGA) is employed to monitor the partial pressures of the dominant species, with particular attention to the oxygen mass channel by mass-to-charge ratio, thereby enabling the determination of the oxygen fraction in the total outgassing rate~\cite{OHanlon2003}. The corresponding normalized oxygen outgassing rates for the main detector materials are summarized in Table~\ref{tab:material_outgassing_swapped}. For each material, a reference specific outgassing rate, \(q_{\mathrm{ref}}\), is defined from the room temperature measurement at \(T_{\mathrm{ref}}\simeq 295~\si{\kelvin}\) after a fixed pump-down time of \(t_{\mathrm{ref}}=72~\mathrm{h}\). The oxygen outgassing rate is normalized either by the exposed surface area or by unit count, as indicated in the Property Unit column.

\begin{table}[htbp]
    \centering
    \caption{Summary of normalized oxygen outgassing rates for key detector materials. The values represent the rates measured at room temperature after 72 hours of pumping, normalized by surface area or unit count.}
    \label{tab:material_outgassing_swapped}
    \begin{tabular}{lcc}
        \toprule
        \\[-8pt]
        \textbf{Material}  & \textbf{Property Unit} & \textbf{\quad Oxygen Outgassing Rate \quad}  \\[1pt]
        &  &[ppb $\cdot$ mol/ h] \\[1pt]
        \toprule
        \\[-8pt]
        Stainless Steel       & \si{cm^{-2}} & $(4.21 \pm 0.16) \times 10^{-4}$ \\[1pt]
        PMT base      & pc (unit)    & $(6.40 \pm 0.66) \times 10^{-1}$ \\[1pt]
        PEEK        & \si{cm^{-2}} & $(2.85 \pm 0.64) \times 10^{-1}$ 
        \\[1pt]
        Kapton       & \si{cm^{-1}} & $(1.49 \pm 0.21) \times 10^{-1}$ 
        \\[1pt]
        PTFE         & \si{cm^{-2}} & $(5.54 \pm 0.30) \times 10^{-1}$
        \\[1pt]
        PTFE(baked)        & \si{cm^{-2}} & $(1.05 \pm 0.77) \times 10^{-1}$ \\[1pt]
        \bottomrule
    \end{tabular}
\end{table}

The values listed in Table~\ref{tab:material_outgassing_swapped} correspond to the oxygen outgassing rates measured at room temperature. Among them, the outgassing rate of stainless steel is significantly lower than that of other materials and can be neglected in the calculations. We measure the pressure change rate in a sealed stainless steel chamber with a volume of approximately 14~L each time pumping is stopped. The outgassing rate is then obtained according to the ideal gas law \(PV = nRT\), with units converted from ppb \(\cdot\) Pa \(\cdot\) L / s to ppb\(\cdot\)mol/h per unit area. Considering the surface areas listed in Table~\ref{tab:detector_outgassing}, the main contributors to the outgassing rate are the reflector materials (PTFE or PEEK) and Kapton. For RELICS-10 and RELICS-50, the evaluation indicates that PTFE provides the dominant contribution to the total oxygen outgassing rate. It was therefore selected for an additional vacuum-baking treatment at \SI{493}{\kelvin} to mitigate its outgassing contribution. After approximately 320 hours of baking, followed by re-exposure to air for about 48 hours, the room-temperature oxygen outgassing rate of PTFE decreased to approximately 0.19 times its original value. The corresponding measured value after baking is included in Table~\ref{tab:material_outgassing_swapped} and is used in the outgassing estimate for future RELICS detectors.

Since the detector operates under cryogenic conditions, the actual outgassing rates of these materials are expected to be lower than their room temperature values. Their temperature dependence is described using an empirical Arrhenius-type relation in a diffusion-limited approximation~\cite{Chiggiato2006}. The temperature effect on the outgassing rate can be expressed as:

\begin{equation}
\begin{aligned}
f(T) &\propto D(T)^{1/2} \\[10pt]
D(T) &= D_0\exp\left(-\frac{E_a}{R T}\right) \\[10pt]
\eta
= \frac{f(T_{\mathrm{low}})}{f(T_{\mathrm{room}})}
&= \exp\left[
-\frac{E_a}{2R}
\left(
\frac{1}{T_{\mathrm{low}}}
-\frac{1}{T_{\mathrm{room}}}
\right)
\right],
\end{aligned}
\label{outgassing_temp}
\end{equation}
where \(f(T)\) is the outgassing rate at temperature \(T\), \(D(T)\) is the diffusion coefficient, \(D_0\) is the pre-exponential factor, \(E_a\) is the activation energy for oxygen transport, \(R\) is the universal gas constant, \(T_{\mathrm{low}}\) is the cryogenic temperature, \(T_{\mathrm{room}}\) is the room temperature, and \(\eta\) is the temperature reduction factor defined as the ratio of the outgassing rate at low temperature to that at room temperature. For PTFE, PEEK, and Kapton, the activation energy values \(E_a\) are taken as 26.3 kJ/mol, 15.8 kJ/mol, and 31 kJ/mol, respectively, based on literature data~\cite{brandrup1999polymer, mergen2003gas, celina2018oxygen}.

\begin{table}[htbp]
    \centering
    \caption{Summary of material quantities for the prototype Run~7, Run~9, and the RELICS experiment. The definition of \(M_i\) is explained in detail in Section~\ref{sec:model_all}. The area column represents the surface area of PTFE or PEEK, while the length column represents the length of the Kapton cables.}
    \label{tab:detector_outgassing}
    \begin{tabular}{lclclc}

        \toprule
        \\[-8pt]
        \textbf{Run} & \textbf{\quad Volume \quad} & 
        \makecell{\textbf{Material}} & 
        \makecell{\textbf{Area} } & 
        \makecell{\textbf{Material}} & 
        \makecell{\textbf{Length} } \\[1pt]
         &  & & [$cm^{2}$] & & [m] \\[1pt]
        \toprule
        \\[-8pt]
        \multirow{4}{*}{Run~7} 
        & \(M_{0}\) & \multirow{4}{*}{PTFE} & 557   & \multirow{12}{*}{Kapton} & 0 \\[1pt]
        & \(M_{1}\) & & 3342 & & 2 \\[1pt]
        & \(M_{4}\) & & 95   & & 0 \\[1pt]
        & \(M_{5}\) & & 230  & & 18 \\[1pt]
        \multirow{3}{*}{Run~9} 
        & \(M_{0}\) & \multirow{3}{*}{PEEK} & 867   & & 0 \\[1pt]
        & \(M_{1}\) & & 1142 & & 2 \\[1pt]
        & \(M_{4}\) & & 414  & & 23 \\[1pt]
        \multirow{3}{*}{RELICS-10} 
        & \(M_{0}\) & \multirow{3}{*}{PTFE} & 1323  & & 0 \\[1pt]
        & \(M_{1}\) & & 1323 & & 12 \\[1pt]
        & \(M_{4}\) & & 755  & & 80 \\[1pt]
        \multirow{3}{*}{RELICS-50} 
        & \(M_{0}\) & \multirow{3}{*}{PTFE} & 4709  & & 0 \\[1pt]
        & \(M_{1}\) & & 4709 & & 64 \\[1pt]
        & \(M_{4}\) & & 2559 & & 288 \\[1pt]
        \bottomrule
    \end{tabular}
\end{table}

An estimate of the oxygen outgassing rate inside the detector is then obtained by combining the room temperature measurements with the amount of each material used in the detector (e.g., PTFE and Kapton), the temperature conditions experienced by each material during operation, and literature values of the activation energy for oxygen transport in the relevant materials. Taking PTFE as an example, its activation energy for oxygen outgassing is reported as 26.3~kJ/mol~\cite{brandrup1999polymer}. At liquid xenon temperatures (\(\sim168\) K) and gaseous xenon temperatures (\(\sim195\) K), the outgassing rate reduction factors due to temperature effects are approximately \(1/58\) and \(1/15.6\) of the room-temperature value, respectively. However, different databases provide significantly varying values of the activation energy for PTFE. For example, PTFE begins to wear by a thermoactivation mechanism with an activation energy of 40--60 ~kJ/mol, while according to the Kissinger method, the activation energy \(E_a\) related to the thermal decomposition kinetics of PTFE was found to be 75.23 ~kJ/mol~\cite{smurugov1992ptfe, seo2000nonisothermal, chu2024mechanistic,dogan2021kinetic}, with discrepancies even reaching orders of magnitude. These deviations may arise from differences in manufacturing processes or from ambiguities in whether the reported activation energies specifically correspond to the transport of oxygen. As a result, the uncertainty in the activation energy parameters for each material remains unconstrained, leading to potentially large and unquantifiable errors in the low-temperature correction factors. Therefore, the discrepancy between the fitted outgassing rates and the expectations derived from room-temperature measurements and subsequent extrapolation to cryogenic conditions, as observed in both Run~7 and Run~9 and summarized in Table~\ref{table:combined_all}, may be explained by this factor.

\subsection{Projection of purification performance for the RELICS experiment}

Based on the validation of the purification model against both the Run~7 and Run~9 prototype data, the model is extended to forecast the long-term LXe purity performance of the future RELICS-10 and RELICS-50 detectors. The projection relies on scaling the validated physical processes to the new detector geometries and operational parameters.

The input parameters for the forecast are determined as follows:
\begin{itemize}
    \item \textbf{Material Outgassing Rates:} The gas load is estimated based on the measured outgassing rates of individual materials and their respective surface areas within each detector volume. Due to the uncertainties inherent in extrapolating from room-temperature measurements to cryogenic conditions, the final outgassing rate is determined by scaling the best-fit outgassing rates from Run~7 according to the surface area. The error includes the statistical uncertainty from the model fitting, the measurement uncertainty of the room-temperature outgassing rates, and the uncertainty in surface area estimation.
    
    \item \textbf{Purifier and Line Efficiency:} The purification efficiency $e$ of the getter and the liquid delivery line sealing efficiency $p$ are adopted from the fitting results of the prototype runs. For the future detectors, the same type of getter will be used to maintain consistently high purification efficiency. The liquid delivery line will employ the reinforced VCR-and-hose structure from Run~9, with the goal of achieving a sealing integrity equal to or greater than that measured in Run~9.
    
    \item \textbf{Gas-Liquid Exchange:} The strength coefficient $h$ of the gas-liquid exchange is obtained from the joint fitting of Run~7 and Run~9. Since the hardware conditions, including the inner and outer vessels and the cryocooler, remained unchanged, the heat leakage is considered to be not significantly different between the two runs. Moreover, the future detectors have no order-of-magnitude difference in scale and will be designed to minimize heat leakage as much as possible. Therefore, this parameter, as determined from the prototype fitting, is extended to the subsequent predictions.
    \item \textbf{Electric Field Effect:} For the purity forecast, the impact of electric field non-uniformity ($\tau_E$) is neglected, focusing solely on the impurity-limited electron lifetime $\tau_{\mathrm{true}}$.
\end{itemize}

Both RELICS-10 and RELICS-50 will retain the overflow chamber structure, the PTFE reflection materials, and the LI-MO circulation mode, with their respective flow rates controlled and monitored by dedicated flow controllers. The target total circulation flow rates are set to \SI{20}{\standardliter\per\minute} for RELICS-10 and \SI{50}{\standardliter\per\minute} for RELICS-50. These possible variations are treated as uncertainties in the projection of the electron lifetime. In terms of detector architecture, both future detectors will employ the overflow chamber design for stable, passive liquid level control. This design continuity means that the fundamental impurity transport mechanisms remain consistent with those characterized in the Run~9 prototype. Therefore, the purification model validated against Run~9 serves as the direct foundation for the purity predictions for RELICS. The key step is to update the model with the specific geometric volume partitions and total xenon masses of the RELICS-10 and RELICS-50 designs. The xenon mass contained within each control volume is also subject to minor variations during actual operation. However, these mass fluctuations do not affect the final purity level. They only slightly influence the rate of increase during the rising phase. These final input parameters and their associated uncertainties are summarized in Table~\ref{table:combined_all}.

\begin{figure}[htbp]
    \centering
    \includegraphics[width=0.99\linewidth]{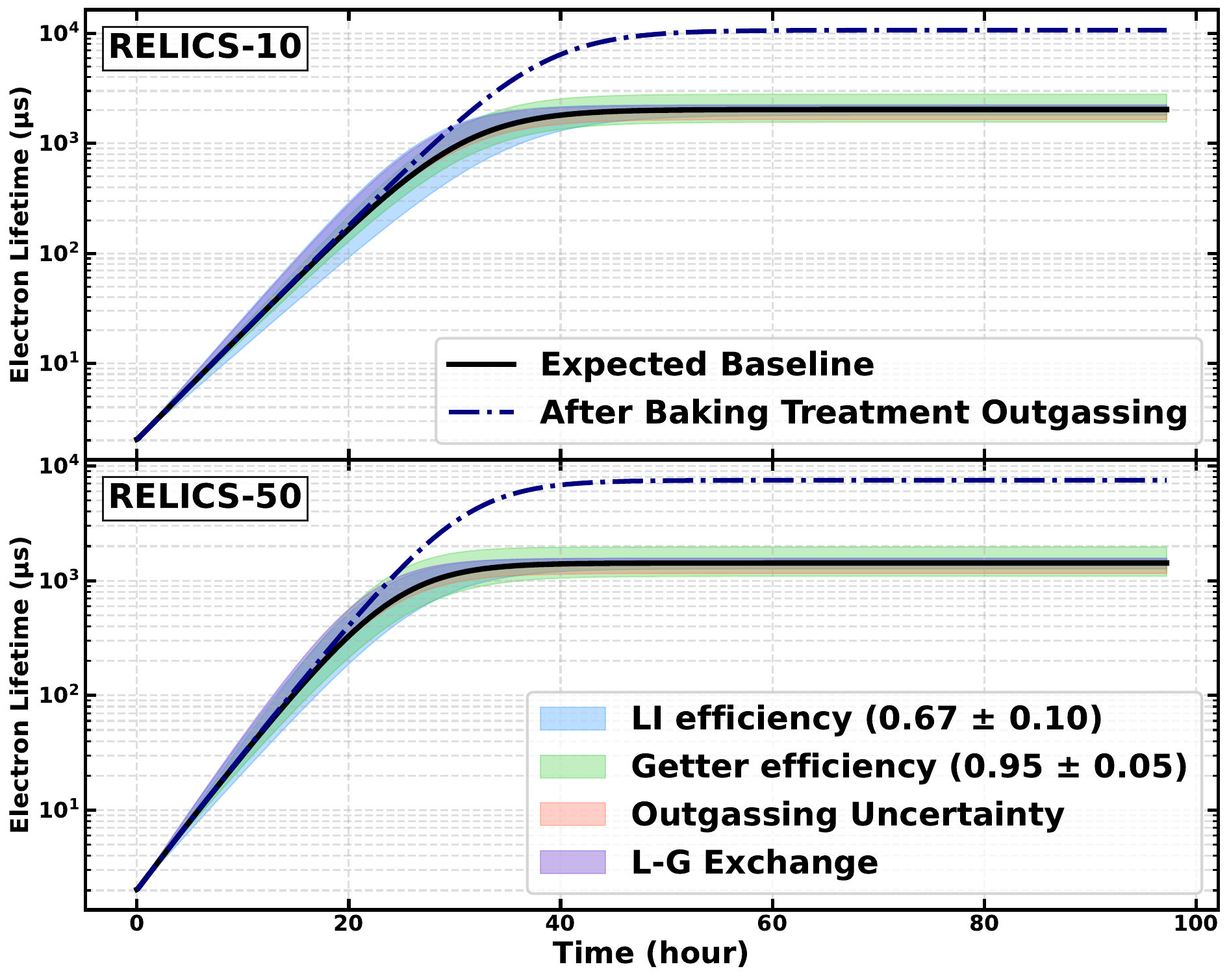}
    \caption{ Prediction results for RELICS-10 and RELICS-50 are presented separately. The baseline curve represents the electron lifetime prediction obtained using the central values of the parameters listed in Table~\ref{table:combined_all}, with the outgassing rates derived from the fitted measurement values. The colored error bands represent the impact of different parameter uncertainties on the predictions. The blue dashed line corresponds to the predicted electron lifetime under the ideal scenario, where the optimal outgassing level is achieved through baking treatment of the materials.}
    \label{fig:relics_prediction}
\end{figure}

Considering the influence of major parameters on purity, the predictions for RELICS-10 and RELICS-50 incorporate both the constrained model parameters and material outgassing rates. The final purity projections are presented in Fig.~\ref{fig:relics_prediction}, while a detailed breakdown of each parameter's individual influence is provided in Table~\ref{table:parameter_sensitivity-combined}. Since the uncertainty in the extrapolation coefficient from room temperature to cryogenic conditions has not yet been clarified, the material outgassing rates used in the projection are derived from fitted values. As mentioned in Section~\ref{subsec:outgassing}, we have identified a method to reduce material outgassing rates. Prior to installation, the material samples are vacuum-baked at \SI{493}{\kelvin}. After approximately 320 hours of baking, these materials are subsequently re-exposed to air for about 48 hours. This treatment reduces the outgassing rate to approximately one-fifth of its original value, with the corresponding result shown as the blue dashed line in the figure. However, this condition is highly sensitive to on-site installation practices. If the exposure time after baking is too long, the outgassing rate may recover to its original level. Therefore, the baking effect must be considered in light of actual installation conditions. This uncertainty has not yet been taken into account, and consequently, no error band is provided.

\begin{table}[htbp]
\centering
\caption{Systematic uncertainty in the projected electron lifetime for RELICS-10 and RELICS-50. The systematic uncertainty in the projected electron lifetime for RELICS-10 and RELICS-50 is evaluated by varying individual parameters and assessing their impact on the final electron lifetime.}
\label{table:parameter_sensitivity-combined}

\begin{tabular}{lcc}
\toprule

\textbf{Parameter} & \multicolumn{2}{c}{\textbf{Electron lifetime}} \\
& \multicolumn{2}{c}{[\si{\milli\second}]} \\

& \textbf{\qquad RELICS-10 \qquad } & \textbf{ \qquad RELICS-50 \qquad} \\
\toprule
\\[-8pt]
$p$  & 1.81 - 2.25 & 1.27 - 1.57 \\

$e$ &  1.56 - 2.80 & 1.16 - 1.97 \\

$f_i$ &  1.65 - 2.03 & 1.16 - 1.43 \\

$h$ &  2.02 - 2.22 & 1.42 - 1.56 \\

$\textbf{Base Lifetime}$ & \textbf{2.03}& \textbf{1.43 }\\
\bottomrule
\end{tabular}
\end{table}

As summarized in Table~\ref{table:combined_all}, the purification model predicts that the RELICS-10 and RELICS-50 detectors are capable of achieving steady-state electron lifetimes of approximately \SI{2.0}{\milli\second} and \SI{1.4}{\milli\second}, respectively. This performance corresponds to reducing the concentration of electronegative impurities in the LXe to the ppb level. Following the electron survival probability approach, we plan to implement a drift electric field of at least \SI{200}{\V\per\cm}, corresponding to a drift velocity of approximately \SI{0.15}{\cm\per\micro\second} in liquid xenon~\cite{albert2017measurement, aprile2014analysis,akerib2014first}. The drift lengths for RELICS-10 and RELICS-50 are approximately \SI{15}{\cm} and \SI{25}{\cm}, respectively. Under these conditions, the electron survival probability $P_z$ for RELICS-10 reaches approximately 95\% and for RELICS-50 approximately 90\%. For the RELICS detectors, the maximum charge loss due to drift does not exceed 10\%, thereby meeting the stringent purity requirements essential for the success of the CE$\nu$NS measurement program. If all internal detector components undergo a baking treatment prior to assembly, with exposure to air minimized as much as possible, and an extended pump-down period is implemented before xenon filling, the initial outgassing load will be further suppressed. Under such high-purity conditions, where the baseline outgassing rates are already very low, the sensitivity analysis  indicates that the uncertainty in the remaining outgassing rate becomes the dominant factor influencing the final electron lifetime. The getter purification efficiency  and the liquid delivery line sealing efficiency rank as the second most influential parameters.

\section{Summary and Discussion}

\label{sec:sum}

Achieving the goal of detecting low-energy CE$\nu$NS signals at a low threshold in the RELICS experiment requires maintaining ultra-high xenon purity to minimize the signal attenuation. Since the purity is influenced by multiple factors, including material outgassing rates, purification efficiency, and flow dynamics, a quantitative purification model is essential to disentangle their individual impacts. To this end, based on a thorough understanding of the long-term dynamic equilibrium and circulation processes of xenon, we established a model by partitioning the detector into distinct control volumes and describing impurity transport between them. By applying this model to data obtained under different purity conditions and operational modes, it is validated against two prototype campaigns: Run~7 and Run~9. To address the poor sealing issue identified in the Run~7 liquid delivery line, an enhanced structural design was adopted in Run~9. This improvement successfully increased the sealing efficiency from approximately 7\% to 67\%. High requirements for electric field uniformity are also established in this process. Concurrently, vacuum-baking treatments of detector materials are expected to suppress the outgassing rates. The RELICS experiment will retain the overflow chamber structure and the LI-MO circulation mode. Integrating all these refinements, the model predicts that the future RELICS-10 and RELICS-50 detectors can achieve electron lifetimes of approximately \SI{2.0}{\milli\second} and \SI{1.4} {\milli\second} level when operating at circulation flow rates of \SI{20}{\liter\per\minute} and \SI{50}{\liter\per\minute}, respectively. These projected purity levels can meet the requirements for the target physics searches.

In addition, differences are observed between the measured outgassing rates and the expectations, which are derived from room-temperature to cryogenic extrapolation. It is currently suspected that this discrepancy is primarily attributed to the temperature coefficient used in the extrapolation. However, significant variations exist in this coefficient across different databases~\cite{brandrup1999polymer,mergen2003gas,celina2018oxygen,seo2000nonisothermal,chu2024mechanistic,smurugov1992ptfe,dogan2021kinetic}, leading to large systematic uncertainties in the estimated outgassing rates at cryogenic temperatures. Using alternative activation energy values from the literature can change the reduction factor significantly. This discrepancy consequently affects the accuracy of future purity predictions. To further constrain this source of uncertainty, the outgassing rates of PTFE, PEEK and Kapton, which are the primary materials used in the detector, will be directly measured under cryogenic conditions via \textit{in-situ} measurements, using temperature conditions representative of the actual detector environment. This approach will provide a direct, material-specific calibration of the temperature-dependent outgassing behavior, eliminating the need for uncertain extrapolations.

\section*{Author Contributions}

All authors contributed to the study conception and design. Material preparation, data collection, analysis, and the first draft of the manuscript were completed by Jia-Chen Yu, Kai-Hang Li, and Jing-Fan Gu. The experimental setup and operation were carried out with the participation of Ling-Feng Xie and Yi-Fei Zhao. This project was led and supervised by Qing Lin and Fei Gao. All authors commented on previous versions of the manuscript. All authors read and approved the final manuscript.

\section*{Declarations}

Conflict of interest: The authors declare that they have no conflict of 
interest.

\section*{Acknowledgments}
We thank Prof. Xiaopeng Zhou from Beihang University for providing the $^{37}$Ar and $^{83\text{m}}\text{Kr}$  sources to calibrate the RELICS prototype.

RELICS is supported by grants from National Key R\&D program from the Ministry of Science and Technology of China (No. 2021YFA1601600), Natural Science Foundation of China (Nos. 12275267, 12521007, 12405129, 12375095, 12250011, 12305121), Ministry of Education of China (No. SRICSPYF-ZY2025028), Beijing Natural Science Foundation (Nos. QY23088, QY25008), Guangzhou Municipal Science and Technology Project (No. 2025A04J5409), Zhejiang Provincial Natural Science Foundation of China (No. LQKWL26A0501), Pengcheng Peacock Program Supporting Research Funding (Class C, No. KQ002778), and the CUHK-Shenzhen University Development Fund (No. UDF01003491).

We acknowledge CNNC Sanmen Nuclear Power Company for hosting RELICS.



\begin{thebibliography}{66}

\bibitem{cevns_theory} D.Z. Freedman, Coherent neutrino nucleus scattering as a probe of the weak neutral current. Phys. Rev. D {\bf 9}, 1389--1392 (1974). \href{https://doi.org/10.1103/PhysRevD.9.1389}{https://doi.org/10.1103/PhysRevD.9.1389}

\bibitem{kopeliovich1974isotopic} V.B. Kopeliovich and L.L. Frankfurt, Isotopic and chiral structure of neutral current. JETP Lett. {\bf 19}, 145--147 (1974).

\bibitem{canas2018future} B.C. Cañas, E.A. Garcés, O.G. Miranda, and A. Parada, Future perspectives for a weak mixing angle measurement in coherent elastic neutrino nucleus scattering experiments. Phys. Lett. B {\bf 784}, 159--162 (2018). \href{https://doi.org/10.1016/j.physletb.2018.07.049}{https://doi.org/10.1016/j.physletb.2018.07.049}

\bibitem{beacom2010diffuse} J.F. Beacom, The diffuse supernova neutrino background. Ann. Rev. Nucl. Part. Sci. {\bf 60}, 439--462 (2010). \href{https://doi.org/10.1146/annurev.nucl.010909.083331}{https://doi.org/10.1146/annurev.nucl.010909.083331}

\bibitem{Coloma:2017egw} P. Coloma, P.B. Denton, M.C. Gonzalez-Garcia, M. Maltoni, and T. Schwetz, Curtailing the dark side in non-standard neutrino interactions. JHEP {\bf 04}, 116 (2017). \href{https://doi.org/10.1007/JHEP04(2017)116}{https://doi.org/10.1007/JHEP04(2017)116}

\bibitem{PhysRevC.95.025801} C.J. Horowitz et al., Neutrino-nucleon scattering in supernova matter from the virial expansion. Phys. Rev. C {\bf 95}, 025801 (2017). \href{https://doi.org/10.1103/PhysRevC.95.025801}{https://doi.org/10.1103/PhysRevC.95.025801}

\bibitem{COHERENT:2017ipa} D. Akimov et al. (COHERENT Collaboration), Observation of coherent elastic neutrino-nucleus scattering. Science {\bf 357}, 1123--1126 (2017). \href{https://doi.org/10.1126/science.aao0990}{https://doi.org/10.1126/science.aao0990}

\bibitem{akimov2021first} D. Akimov et al. (COHERENT Collaboration), First measurement of coherent elastic neutrino-nucleus scattering on argon. Phys. Rev. Lett. {\bf 126}, 012002 (2021). \href{https://doi.org/10.1103/PhysRevLett.126.012002}{https://doi.org/10.1103/PhysRevLett.126.012002}

\bibitem{adamski2025evidence} S. Adamski et al. (COHERENT Collaboration), Evidence of coherent elastic neutrino-nucleus scattering with COHERENT's germanium array. Phys. Rev. Lett. {\bf 134}, 231801 (2025). \href{https://doi.org/10.1103/PhysRevLett.134.231801}{https://doi.org/10.1103/PhysRevLett.134.231801}

\bibitem{PhysRevLett.133.191002} E. Aprile et al. (XENON Collaboration), First indication of solar $^{8}\mathrm{B}$ neutrinos via coherent elastic neutrino-nucleus scattering with XENONnT. Phys. Rev. Lett. {\bf 133}, 191002 (2024). \href{https://doi.org/10.1103/PhysRevLett.133.191002}{https://doi.org/10.1103/PhysRevLett.133.191002}

\bibitem{PhysRevLett.133.191001} Z. Bo et al. (PandaX Collaboration), First indication of solar $^{8}\mathrm{B}$ neutrinos through coherent elastic neutrino-nucleus scattering in PandaX-4T. Phys. Rev. Lett. {\bf 133}, 191001 (2024). \href{https://doi.org/10.1103/PhysRevLett.133.191001}{https://doi.org/10.1103/PhysRevLett.133.191001}

\bibitem{Ackermann:2025} N. Ackermann et al., Direct observation of coherent elastic antineutrino–nucleus scattering. Nature {\bf 643}, 1229--1233 (2025). \href{https://doi.org/10.1038/s41586-025-09322-2}{https://doi.org/10.1038/s41586-025-09322-2}

\bibitem{tpc} E. Aprile and T. Doke, Liquid xenon detectors for particle physics and astrophysics. Rev. Mod. Phys. {\bf 82}, 2053--2097 (2010). \href{https://doi.org/10.1103/RevModPhys.82.2053}{https://doi.org/10.1103/RevModPhys.82.2053}

\bibitem{cai2024reactor} C. Cai et al. (RELICS Collaboration), Reactor neutrino liquid xenon coherent elastic scattering experiment. Phys. Rev. D {\bf 110}, 072011 (2024). \href{https://doi.org/10.1103/PhysRevD.110.072011}{https://doi.org/10.1103/PhysRevD.110.072011}

\bibitem{qian2019physics} X. Qian and J.-C. Peng, Physics with reactor neutrinos. Rept. Prog. Phys. {\bf 82}, 036201 (2019). \href{https://doi.org/10.1088/1361-6633/aae881}{https://doi.org/10.1088/1361-6633/aae881}

\bibitem{xie2026development} L. Xie et al., Development of a dual-phase xenon time projection chamber prototype for the RELICS experiment. Eur. Phys. J. C {\bf 86}, 348 (2026). \href{https://doi.org/10.1140/epjc/s10052-026-15529-y}{https://doi.org/10.1140/epjc/s10052-026-15529-y}

\bibitem{yang2026design} J. Yang et al. (RELICS Collaboration), Design and characterization of a photosensor system for the RELICS experiment. JINST {\bf 21}, P02039 (2026). \href{https://doi.org/10.1088/1748-0221/21/02/P02039}{https://doi.org/10.1088/1748-0221/21/02/P02039}

\bibitem{guo2026preparation} X.-N. Guo et al., Preparation and measurement of an $^{37}\textrm{Ar}$ source for liquid xenon detector calibration. Nucl. Sci. Tech. {\bf 37}, 14 (2026). \href{https://doi.org/10.1007/s41365-025-01818-w}{https://doi.org/10.1007/s41365-025-01818-w}

\bibitem{elife_02} G. Bakale, U. Sowada, and W.F. Schmidt, Effect of an electric field on electron attachment to sulfur hexafluoride, nitrous oxide, and molecular oxygen in liquid argon and xenon. J. Phys. Chem. {\bf 80}, 2556--2559 (1976). \href{https://doi.org/10.1021/j100564a006}{https://doi.org/10.1021/j100564a006}

\bibitem{zhang2022rb} D. Zhang et al., Rb83/Kr83m production and cross-section measurement with 3.4 MeV and 20 MeV proton beams. Phys. Rev. C {\bf 105}, 014604 (2022). \href{https://doi.org/10.1103/PhysRevC.105.014604}{https://doi.org/10.1103/PhysRevC.105.014604}

\bibitem{be2013table} M.M. B{\'e}, V. Chist{\'e}, C. Dulieu, E. Browne, V. Chechev, N. Kuzmenko, R. Helmer, A. Nichols, E. Sch{\"o}nfeld, and R. Dersch, Table of radionuclides, Monographie BIPM-5, vol. 7. Bureau International des Poids et Mesures, S{\`e}vres (2013).

\bibitem{Plante:2022khm} G. Plante, E. Aprile, J. Howlett, and Y. Zhang, Liquid-phase purification for multi-tonne xenon detectors. Eur. Phys. J. C {\bf 82}, 860 (2022). \href{https://doi.org/10.1140/epjc/s10052-022-10832-w}{https://doi.org/10.1140/epjc/s10052-022-10832-w}

\bibitem{o2henry} R.A. Pierotti, The solubility of gases in liquids. J. Phys. Chem. {\bf 67}, 1840--1845 (1963). \href{https://doi.org/10.1021/j100803a024}{https://doi.org/10.1021/j100803a024}



\bibitem{foreman2013emcee} D. Foreman-Mackey, D.W. Hogg, D. Lang, and J. Goodman, emcee: The MCMC Hammer. Publ. Astron. Soc. Pac. {\bf 125}, 306 (2013). \href{https://doi.org/10.1086/670067}{https://doi.org/10.1086/670067}





\bibitem{foreman2019emcee} D. Foreman-Mackey, W.M. Farr, M. Sinha, A.M. Archibald, D.W. Hogg, J.S. Sanders, J. Zuntz, P.K.G. Williams, A.R.J. Nelson, M. de Val-Borro, et al., emcee v3: A Python ensemble sampling toolkit for affine-invariant MCMC. J. Open Source Softw. {\bf 4}, 1864 (2019). \href{https://doi.org/10.21105/joss.01864}{https://doi.org/10.21105/joss.01864}




\bibitem{CHIB20013569} S. Chib, Markov chain Monte Carlo methods: Computation and inference, in Handbook of Econometrics, Vol. 5, J.J. Heckman and E. Leamer, eds. (Elsevier, 2001), pp. 3569--3649. \href{https://doi.org/10.1016/S1573-4412(01)05010-3}{https://doi.org/10.1016/S1573-4412(01)05010-3}



\bibitem{mackay2003information} D.J.C. MacKay, Information Theory, Inference and Learning Algorithms (Cambridge University Press, Cambridge, 2003).


\bibitem{aprile2023low} E. Aprile et al. (XENON Collaboration), Low-energy calibration of XENON1T with an internal $^{37}$Ar source. Eur. Phys. J. C {\bf 83}, 542 (2023). \href{https://doi.org/10.1140/epjc/s10052-023-11512-z}{https://doi.org/10.1140/epjc/s10052-023-11512-z}




\bibitem{szydagis2025review} M. Szydagis, J. Balajthy, G.A. Block, J.P. Brodsky, E. Brown, J.E. Cutter, S.J. Farrell, J. Huang, A.C. Kamaha, E.S. Kozlova, et al., A review of NEST models for liquid xenon and an exhaustive comparison with other approaches. Front. Detector Sci. Technol. {\bf 2}, 1480975 (2025). \href{https://doi.org/10.3389/fdest.2024.1480975}{https://doi.org/10.3389/fdest.2024.1480975}





\bibitem{akimov2014experimental} D.Y. Akimov et al., Experimental study of ionization yield of liquid xenon for electron recoils in the energy range 2.8--80 keV. JINST {\bf 9}, P11014 (2014). \href{https://doi.org/10.1088/1748-0221/9/11/P11014}{https://doi.org/10.1088/1748-0221/9/11/P11014}



\bibitem{OHanlon2003} J.F. O'Hanlon, A User's Guide to Vacuum Technology, 3rd ed. (John Wiley \& Sons, Hoboken, NJ, 2003). \href{https://doi.org/10.1002/0471467162}{https://doi.org/10.1002/0471467162}




\bibitem{Fedchak2021} J.A. Fedchak, J.K. Scherschligt, S. Avdiaj, D.S. Barker, S.P. Eckel, B. Bowers, S. O'Connell, and P. Henderson, Outgassing rate comparison of seven geometrically similar vacuum chambers of different materials and heat treatments. J. Vac. Sci. Tech. B {\bf 39}, 024201 (2021). \href{https://doi.org/10.1116/6.0000657}{https://doi.org/10.1116/6.0000657}



\bibitem{Chiggiato2006} P. Chiggiato, Outgassing properties of vacuum materials for particle accelerators (2020). \href{https://doi.org/10.48550/arXiv.2006.07124}{https://doi.org/10.48550/arXiv.2006.07124}


\bibitem{brandrup1999polymer} J. Brandrup, E.H. Immergut, E.A. Grulke, A. Abe, and D.R. Bloch, Polymer Handbook, Vol. 89 (Wiley, New York, 1999).

\bibitem{mergen2003gas} G. Mergen, Gas permeation properties of poly (arylene ether ketone) and its mixed matrix membranes with polypyrrole, Master's thesis, Middle East Technical University (2003).

\bibitem{celina2018oxygen} M.C. Celina and A. Quintana, Oxygen diffusivity and permeation through polymers at elevated temperature. Polymer {\bf 150}, 326--342 (2018). \href{https://doi.org/10.1016/j.polymer.2018.06.047}{https://doi.org/10.1016/j.polymer.2018.06.047}


\bibitem{smurugov1992ptfe} V.A. Smurugov, A.I. Senatrev, V.G. Savkin, V.V. Biran, and A.I. Sviridyonok, On PTFE transfer and thermoactivation mechanism of wear. Wear {\bf 158}, 61--69 (1992). \href{https://doi.org/10.1016/0043-1648(92)90030-C}{https://doi.org/10.1016/0043-1648(92)90030-C}

\bibitem{seo2000nonisothermal} Y. Seo, Nonisothermal crystallization kinetics of polytetrafluoroethylene. Polym. Eng. Sci. {\bf 40}, 1293--1297 (2000). \href{https://doi.org/10.1002/pen.11257}{https://doi.org/10.1002/pen.11257}


\bibitem{chu2024mechanistic} C. Chu, L.L. Ma, H. Alawi, W. Ma, Y. Zhu, J. Sun, Y. Lu, Y. Xue, and G. Chen, Mechanistic exploration of polytetrafluoroethylene thermal plasma gasification through multiscale simulation coupled with experimental validation. Nat. Commun. {\bf 15}, 1654 (2024). \href{https://doi.org/10.1038/s41467-024-45077-6}{https://doi.org/10.1038/s41467-024-45077-6}





\bibitem{dogan2021kinetic}
F. Dogan, A kinetic evaluation for phenothiazine based copolymers. Hittite J. Sci. Eng. {\bf 8}, 63--70 (2021). \href{https://doi.org/10.17350/HJSE19030000214}{https://doi.org/10.17350/HJSE19030000214}



\bibitem{albert2017measurement} J.B. Albert et al. (EXO-200 Collaboration), Measurement of the drift velocity and transverse diffusion of electrons in liquid xenon with the EXO-200 detector. Phys. Rev. C {\bf 95}, 025502 (2017). \href{https://doi.org/10.1103/PhysRevC.95.025502}{https://doi.org/10.1103/PhysRevC.95.025502}


\bibitem{aprile2014analysis} E. Aprile et al. (XENON100 Collaboration), Analysis of the XENON100 dark matter search data. Astropart. Phys. {\bf 54}, 11--24 (2014). \href{https://doi.org/10.1016/j.astropartphys.2013.10.002}{https://doi.org/10.1016/j.astropartphys.2013.10.002}



\bibitem{akerib2014first} D.S. Akerib et al. (LUX Collaboration), First results from the LUX dark matter experiment at the Sanford Underground Research Facility. Phys. Rev. Lett. {\bf 112}, 091303 (2014). \href{https://doi.org/10.1103/PhysRevLett.112.091303}{https://doi.org/10.1103/PhysRevLett.112.091303}

\end{thebibliography}

\end{document}